\documentclass[printer]{aa}
\usepackage{graphicx}
\usepackage{natbib}
\bibpunct{(}{)}{;}{a}{}{,} % to follow the A&A style

\def\kms{km~s$^{-1}$}
\def\tef{\textit{T}_{\text{eff}}}
\def\logg{\text{log}(\textit{g})}
\def\mh{[\text{M}/\text{H}]}
\def\fe2h{[\text{FeII}/\text{H}]}
\def\feh{[\text{Fe}/\text{H}]}
\def\mg1fe1{[\ion{Mg}{I}/\ion{Fe}{I}]}
\def\mgfe{[\text{Mg}/\text{Fe}]}
\def\alffe{[\alpha/\text{Fe}]}
\def\snr{S/N}
\def\vr{\text{V}_R}
\def\vf{\text{V}_{\phi}}
\def\vz{\text{V}_Z}

\def\meanvr{<V_{R}>}
\def\meanvf{<V_{\phi}>}
\def\meanvz{<V_{Z}>}

\def\sigr{\sigma_{R}}
\def\sigf{\sigma_{\phi}}
\def\sigz{\sigma_{Z}}
\def\vrad{V_{\text{rad}}}

\def\gradmgf{\nabla\text{V}_{\phi}/\nabla[\text{Mg}/\text{Fe}]}

\def\gradfef{\nabla\text{V}_{\phi}/\nabla[\text{Fe}/\text{H}]}

\def\gradvphiz{\nabla\text{V}_{\phi}/\nabla Z}

\def\kmsdex{km~s$^{-1}$~dex$^{-1}$}
\def\kmskpc{km~s$^{-1}$~kpc$^{-1}$}

\begin{document}

\title{The Gaia-ESO Survey:\\ New constraints on the Galactic disc velocity dispersion 
and its chemical dependencies \thanks{Based on observations collected with the 
FLAMES spectrograph at the VLT/UT2 telescope (Paranal Observatory, ESO, Chile), 
for the Gaia-ESO Large Public Survey, programme 188.B-3002}}

\titlerunning{The Galactic disc velocity dispersion}
\authorrunning{G. Guiglion and collaborators}

\author{G. Guiglion \inst{1}, 
A. Recio-Blanco \inst{1}, 
P. de Laverny \inst{1}, 
G. Kordopatis \inst{2}, 
V. Hill \inst{1}, 
\v S. Mikolaitis \inst{1}, 
I. Minchev \inst{2}, 
C. Chiappini \inst{2}, 
R. F. G. Wyse  \inst{3}, 
G. Gilmore \inst{4}, 
S. Randich \inst{5}, 
S. Feltzing \inst{6}, 
T. Bensby \inst{6}, 
E. Flaccomio \inst{7}, 
S.~E. Koposov \inst{4, 8}
E. Pancino \inst{9}, 
A. Bayo \inst{10}, 
M.~T. Costado \inst{11}, 
E. Franciosini \inst{5}, 
A. Hourihane \inst{4}, 
P. Jofr\'e \inst{4}, 
C. Lardo \inst{12}, 
J. Lewis \inst{4}, 
K. Lind \inst{13}, 
L. Magrini \inst{5}, 
L. Morbidelli \inst{5}, 
G.~G. Sacco \inst{5}, 
G. Ruchti \inst{6}, 
C.~C. Worley \inst{4}, 
S. Zaggia \inst{14}}

\institute{Laboratoire Lagrange, Universit\'e C\^ote d'Azur, Observatoire de la C\^ote d'Azur, CNRS, 
Blvd de l'Observatoire, CS 34229, 06304 Nice cedex 4, France
\and Leibniz-Institut für Astrophysik Potsdam (AIP) An der Sternwarte 16, D-14482 Potsdam, Germany 
\and Physics and Astronomy Department, Johns Hopkins University, 3400 North Charles Street, Baltimore, MD 21218, USA
\and Institute of Astronomy, University of Cambridge, Madingley Road, Cambridge CB3 0HA, United Kingdom
 \and INAF - Osservatorio Astrofisico di Arcetri, Largo E. Fermi 5, 50125, Florence, Italy
\and Lund Observatory, Department of Astronomy and Theoretical Physics, Box 43, SE-221 00 Lund, Sweden 
\and INAF - Osservatorio Astronomico di Palermo, Piazza del Parlamento 1, 90134, Palermo, Italy 
\and Moscow MV Lomonosov State University, Sternberg Astronomical Institute, Moscow 119992, Russia 
\and INAF - Osservatorio Astronomico di Bologna, via Ranzani 1, 40127, Bologna, Italy 
ASI Science Data Center, Via del Politecnico SNC, 00133 Roma, Italy
\and  Instituto de F\'isica y Astronomi\'ia, Universidad de Valparai\'iso, Chile
\and Instituto de Astrof\'{i}sica de Andaluc\'{i}a-CSIC, Apdo. 3004, 18080, Granada, Spain 
\and Astrophysics Research Institute, Liverpool John Moores University, 146 Brownlow Hill, Liverpool L3 5RF, United Kingdom 
\and Department of Physics and Astronomy, Uppsala University, Box 516, SE-751 20 Uppsala, Sweden 
\and INAF - Padova Observatory, Vicolo dell'Osservatorio 5, 35122 Padova, Italy}

\date{Received 13/02/2015; accepted 26/08/2015}

\abstract{Understanding the history and the evolution of the Milky Way is 
one of the main goals of modern astrophysics. In particular, the formation of the Galactic 
disc is a key problem of Galactic archaeology.}{We study the velocity dispersion behaviour of Galactic disc 
stars as a function of the $\mgfe$ ratio, which for small metallicity 
bins can be used as a proxy of relative age. This key relation is essential to 
constrain the formation mechanisms of the disc stellar populations as well as the cooling and 
settling processes.}{We used the recommended parameters and chemical abundances 
of $7\,800$ FGK Milky Way field stars from the second internal data release 
of the Gaia-ESO spectroscopic Survey. These stars were observed with the GIRAFFE 
spectrograph (HR10 and HR21 setups), and cover a large spatial volume in the 
intervals $6<R<10\,$kpc and $|Z|<2\,$kpc. Based on a chemical criterion, 
we separated the thin- from the thick-disc sequence in the $\mgfe\,v.s\,\feh$ 
plane.}{From analysing the Galactocentric velocity of
the stars for the thin 
disc, we find  a weak positive correlation between $\vf$ and $\feh$ that is due to  
 a slowly rotating $\feh$-poor tail. For the thick disc stars, a strong correlation 
with $\feh$ and $\mgfe$ is established. In addition, we have detected an inversion 
of the velocity dispersion trends with $\mgfe$ for thick-disc stars with 
$\feh<-0.10\,$dex and $\mgfe>+0.20\,$dex for the radial component. First, the velocity dispersion increases 
with $\mgfe$ at all $\feh$ ratios for the thin-disc stars, and then it decreases for the 
thick-disc population at the highest $\mgfe$ abundances. Similar trends are observed for several 
bins of $\mgfe$ within the errors for the azimuthal velocity dispersion, while a continuous increase with 
$\mgfe$ is observed for the vertical velocity dispersion. The velocity dispersion decrease agrees with previous 
measurements of the RAVE survey, although it is observed here for a greater metallicity interval and a larger 
spatial volume.}{Thanks to the 
Gaia-ESO Survey data, we confirm the existence of $\mgfe$-rich thick-disc stars 
with cool kinematics in the generally turbulent context of the primitive Galactic 
disc. This is discussed in the framework of the different disc formation and 
evolution scenarios.}


\maketitle

\section{Introduction}
During the past decades, strong observational efforts have been made to 
understand the formation and evolution of the Milky Way (MW). To identify and characterise 
the Galactic structures, we need to study two main physical properties: the stellar chemistry 
and the kinematics. The photospheric elemental composition of low-mass 
stars reflects the chemical evolution history of the stellar populations, while the 3D velocities 
are the fossil records of the stellar motions inside the Galaxy. Thus, massive 
spectroscopic surveys are the best tools for distinguishing stellar populations. 

The first massive and low-resolution surveys such as RAVE \citep{RAVE_2006} 
and SEGUE \citep{SEGUE_2009} provided pioneer results with excellent statistics, 
but more detailed chemical information can be obtained from surveys
with a higher resolution, such as 
the Gaia-ESO (GES, \citealt{ges_gilmore_2012}) or APOGEE \citep{allende_2008}.

The existence of two separated thin and thick discs is still a matter of 
debate (see \citealt{bovy_2012_no_thick}) in a context where several scenarios 
are proposed to explain the formation of the Galactic disc stellar populations. 
The vertical heating of a pre-existing disc by a minor merger was proposed by several 
studies \citep{hernquist_1989, quinn_1993, walker_1996, villalobos_2008}. \citet{abadi_2003} 
argued that the thick disc is composed mostly of accreted stars from mergers of dwarf 
galaxies, while the coalescence of a gas-rich satellite forming a thick disc in situ was also 
suggested by \citet{brook_2004, brook_2007}.

More recently, radial migration \citep{sellwood_binney_2002}, driven by the 
co-rotation (churning) or the \mbox{Lindblad} (blurring) resonances with the disc spiral 
structure, has been evoked to explain the emergence of the thick disc. Nevertheless, 
several authors (e.g. \citealt{minchev_2012, martig_2014, veraciro_2014}) used numerical 
models to propose that migration in the sense of churning does not significantly contribute 
to disc thickening.

On the other hand, \citet{bournaud_2009} showed that stellar 
scattering by massive clumps could explain the creation of the
thick disc through a purely internal 
mechanism. Finally, \citet{Haywood_2013} proposed that the early epoch of the disc was driven 
by an intense star formation, which was quantified by \citet{Lehnert_2014} and 
\citet{snaith_2015}. All these proposed scenarios can be constrained by studying the chemo-dynamical 
relations followed by the stellar populations of the disc in a large spatial volume and with robust 
statistics.

To understand the dynamical heating of disc stars, one can investigate the 
age-Velocity-Dispersion (AVD) relation in the very close Solar neighbourhood, which was first studied 
by \citet{wielen_1974, wielen_1975}, who used one thousand stars within $20\,$pc from the Sun. 
These works highlighted the fact that old stars have a higher velocity dispersion. 
\citet{wielen_1977} showed that this interesting behaviour could be explained by the 
diffusion of the orbits due to interactions with the local fluctuations of the 
gravitational field. \citet{carlberg_1985} provided a new AVD relation for 
$\text{about }500$~F-type stars, showing a flat behaviour with age for stars older than 
$6\,$Gyr. Based on the dwarf sample of the Geneva-Copenhagen Survey, 
\citet{nordstrom_2004} found that the AVD relation of the young disc (younger than 
$7\,$Gyr) follows a power law. This was later revised by \citet{seabroke_2007}, however,
who showed that no power law is needed to explain the shape of the relation. 
Finally, \citet{sharma_2014} constrained the thin-disc AVD relation as a power law with RAVE data by 
fitting a kinematic model to the observations.

All these studies are based on small samples with well-constrained distances (Hipparcos volume) or on 
larger samples with a low or intermediate spectral resolution. However, determining precise 
ages for a larger star sample is quite challenging because we lack good distance 
measurements and precise fundamental parameter determinations. For the youngest 
and oldest stars, moreover, the isochrones are crowded in these regimes, and it is harder to derive precise ages. 
For these reasons, the $\alffe$ ratio has been suggested as an age proxy by several authors 
\citep{matteucci_2001, bovy_2012_spatial, minchev_chemo_2014}. Indeed, in the early epoch of the Galaxy, 
the type Ia supernovae (SNIa) ejecta rapidly enriched the interstellar medium and fixed the $\alffe$ ratio at a plateau 
value. Because they evolve on a long timescale, SNIa generate more iron, 
which leads to a decline of the $\alffe$ ratio with increasing $\feh$ and forms a typical break 
(or knee). Interestingly, \citet{Haywood_2013} and \citet{bergemann_2014} found a tight correlation 
between the stellar ages and the $\mgfe$ ratio. We note that \citet{Lee_2011a} observed 
a steeper increase of the velocity dispersion as a function of the $\mgfe$ ratio 
in the thick disc than in the thin component.

More recently, a relation between the velocity dispersion and the $\mgfe$ 
ratio, used as an age proxy, has been presented by \citet{minchev_chemo_2014}. They 
derived a chemo-kinematical relation for the Galactic disc for
$\text{about}5\,000$~giant stars of the Solar suburb ($|Z|<0.6\,$kpc and $7<R<9\,$kpc) 
from the RAVE DR4 survey \citep{kordopatis_2013}. They found a surprising behaviour 
for the most Fe-poor and Mg-rich stars ($\feh<-0.8\,$dex and $\mgfe>+0.4\,$dex): a strong 
decrease of the velocity dispersion is observed while it is expected to increase 
because these stars should correspond to the oldest population of the Galactic disc. We 
note that they also observed this decrease in the SEGUE G-dwarf sample ($R=2000$) from 
\citet{Lee_2011a} and the $\alffe$ ratios, instead of $\mgfe$. In our study, we focus 
on the results obtained with RAVE because they have a higher resolution and were obtained using the $\mgfe$ ratio. 
By comparing their results with numerical simulations, \citet{minchev_chemo_2014} concluded that this 
unexpected behaviour is due to 
perturbations by massive mergers on the outer part of the disc in the early epoch of our Galaxy, 
triggering a strong radial migration process that redistributed stars form the inner disc into the Solar 
suburb on near-circular orbits and lower velocity dispersion.

To extend this last study, we propose here to analyse the behaviour of the 
velocity dispersion of the Galactic disc as a function of its $\mgfe$ ratio with GES data. 
The observations were made with the VLT/UT2 of the European Southern Observatory. 
Our analysis is based on the data from the 
second internal data release (iDR2), which is composed of $\text{about
}8\,000$~FGK dwarf and giant stars. 
We here have the opportunity of probing the Galactic disc more
deeply because GES observes fainter 
targets farther away from the Solar neighbourhood. Moreover, we have a higher precision 
on the atmospheric parameters and chemical abundances thanks to the higher spectral resolution 
and wavelength coverage than was possible with RAVE.

The paper is organised as follows: we first present our stellar sample and chemically 
define the Galactic discs (Sect.~\ref{stellar_sample}), then we detail our method for 
the velocity dispersion derivation (Sect.~\ref{deriv_veloc_disp}). Section~\ref{re_sults} 
is devoted to presenting our results, and we compare our chemo-kinematical relation with the 
one of the RAVE survey in Sect.~\ref{dis_cussion}. Finally, we discuss the constraints 
provided by this study on the different scenarios of the Galactic disc formation in 
Sect.~\ref{impli_cations} and then conclude and summarise our work in Sect.~\ref{con_clusion}.

\section{Stellar sample and characterisation of the Galactic disc}\label{stellar_sample}

Our work is based on the second internal data release (iDR2) 
of the Gaia-ESO spectroscopic Survey \citep{ges_gilmore_2012}. 
We chose a sample of FGK Milky 
Way field stars observed by the GIRAFFE spectrograph in both 
HR10 \mbox{($\text{R}\thicksim19\,800$)} and HR21 
\mbox{($\text{R}\thicksim16\,200$)} setups. In particular, we 
selected stars with available radial velocity $\vrad$, 
effective temperature $\tef$, surface gravity $\logg$, 
global metallicity $\mh$\footnote{\mh~corresponds 
to the proportion of all elements heavier than helium}, and 
$\feh$ and $\mgfe$\footnote{$[\text{Mg}/\text{Fe}]=
\log{\left[\frac{N(\text{Mg})}{N(\text{Fe})}\right]_{\bigstar}}-
~\log{\left[\frac{N(\text{Mg})}{N(\text{Fe})}\right]_{\bigodot}}$} 
abundance ratios; this resulted in a sample of $7\,800$ stars.

We recall that for iDR2, the radial velocity 
$\vrad$ was determined in a first step (complete procedure described 
in Koposov et al., in preparation), with a median error of 
$\text{about }0.3\,$\kms. Then, the fundamental parameters $\tef$, 
$\logg$, $\mh$ and $\alffe$ were computed in three different independent procedures 
(see a short description in \citealt{ges_disc_recio_2014}). The median 
dispersions among these independent measures for our sample are 
$41\,\text{K}$, $0.09\,$dex, and $0.05\,$dex for $\tef$, $\logg,$ 
and $\mh$, respectively. We selected stars with a minimum 
signal-to-noise ratio ($\snr$) equal to 10, leading to distributions 
with a median value of 23 in HR10 and 50 in HR21.

From the atmospheric parameters, $\alpha$ and iron-peak 
elemental abundances were computed by combining 
the determinations of three distinct methods (see 
Recio-Blanco et al., in preparation). To chemically characterise the Galactic disc, we 
focus in the following on the recommended magnesium and iron abundances. 
We adopted the definition \mbox{$\mgfe = \mg1fe1$}, rejecting the
$\ion{Fe}{II}$ abundances that are measured from a single spectral line. In 
the adopted sample, the typical median dispersions of the three 
independent measures are 0.06 and $0.04\,$dex for $\feh$ and $\mgfe$, 
respectively.

To determine the stellar kinematics, we followed 
the procedure described in \citet{kordopatis_2011} and 
\citet{ges_disc_recio_2014}. We recall that we derived the 
absolute magnitudes and then the line-of-sight distances, 
projecting $\tef$, $\logg$ and 
$\mh$ on a set of Yonsei-Yale isochrones 
\citep{y2_demarque_2002}. The three-dimensional Galactic 
coordinates of the stars were then calculated. Finally, 
the Galactocentric radial, azimuthal, and vertical velocities 
($\vr$, $\vf$, and $\vz$, respectively) 
were derived for all the stars with available proper 
motions from the PPMXL catalogue \citep{roeser_2010_ppmxl}. 
We adopted the Local Standard of Rest (LSR) 
$(U, V, W)_{\bigodot}=(11.1, 12.24, 7.25)\,$\kms~of 
\citet{schonrich_2010}. The LSR is presumed to be on a 
circular orbit with an azimuthal velocity $V_{\text{c}}=220\,$\kms.
For the two components of the proper motion $\mu_{\alpha}$ and 
$\mu_{\delta}$, the typical errors are \mbox{$\text{about }7$ mas/yr}. 
The resulting median errors on the whole sample are 52, 39 and 
$52\,$\kms~for $\vr$, $\vf$, and $\vz$, respectively. Finally, our 
\emph{main} sample is composed of $6\,800$ stars with fundamental 
parameters, $\mgfe$ abundances, and 3D kinematics. 

\begin{figure}
\centering
\includegraphics[width=1.0\linewidth]{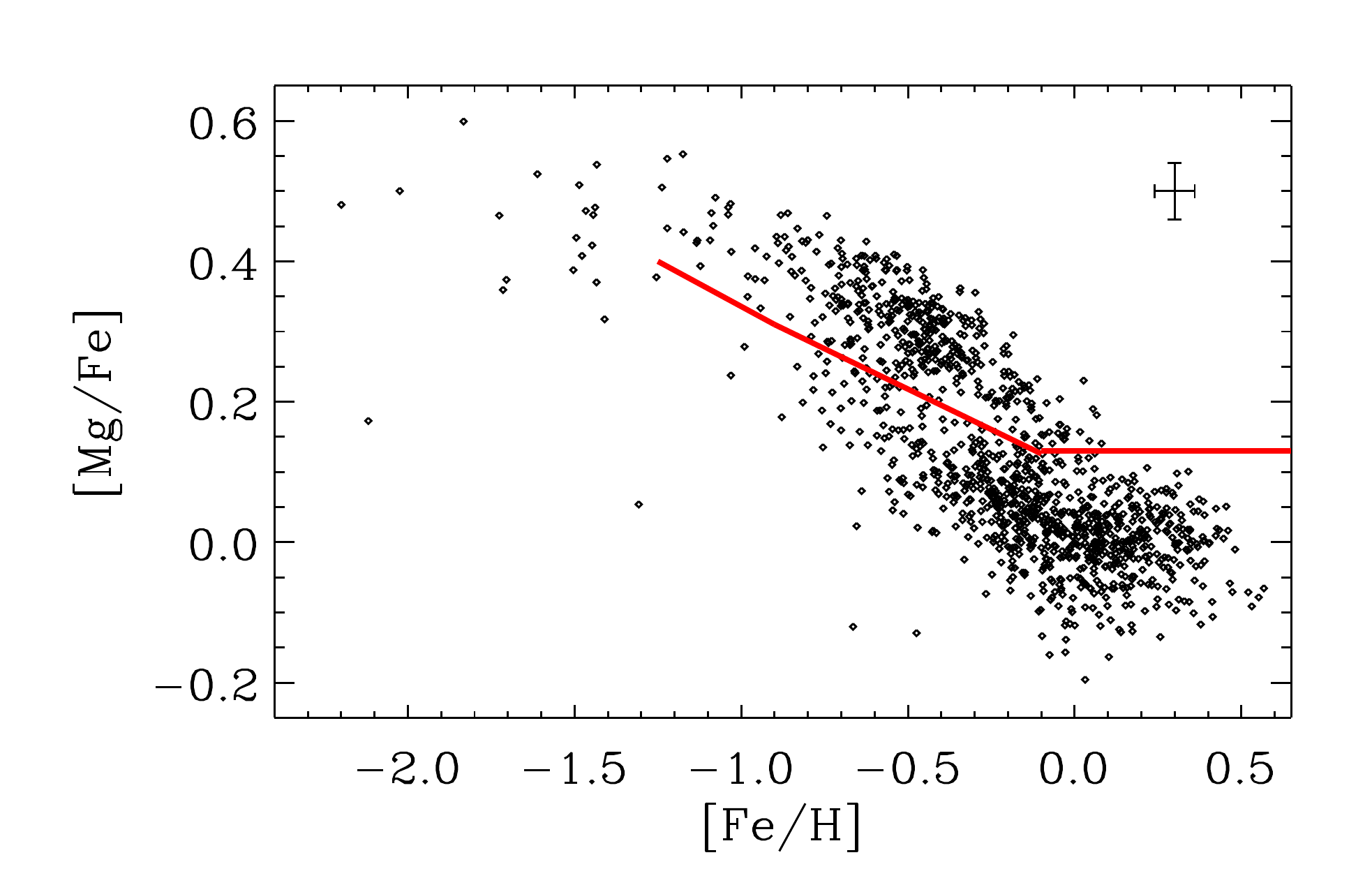}
\caption{\label{thin_thick_transition}$\mgfe$ as a 
function of $\feh$ for the $1\,230$ iDR2 stars with $\snr \ge 40$ 
($16\%$ of the selected sample). The error bars show the typical 
median dispersion. The red solid line represents our division 
between the thin- (below) and the thick- (above) disc stars.}
\end{figure}

We have explored the existence of a chemical gap between 
the thin and the thick disc populations in the $\mgfe\,vs.\,\feh$ plane. 
This gap was first detected in high spectral resolution samples 
in the close Solar neighbourhood by several authors (e.g. \citealt{fuhrmann_1998, 
feltzing_2011, bensby_2003}). It 
has been confirmed more recently by \citet{Adibekyan2011} and 
\citet{bensby_2014} still in the close Solar neighbourhood and beyond 
thanks to GES iDR1 \citep{ges_disc_recio_2014, Sarunas} and APOGEE 
\citep{nidever_2014_apogee} data. To define this gap, we followed the 
same procedure as in \citet{ges_disc_recio_2014} and \citet{Sarunas}. 
We selected every star with $\snr\ge40$, ($1\,230$ stars corresponding 
to $16\%$ of the selected sample) that we decomposed into nine $\feh$ bins 
from -1.50 to $+0.50\,$dex. Then, we searched the minimum count of each bin 
in the corresponding $\mgfe$ distribution. We obtained a quasi-linear 
separation from $\feh=-1.25$ to $-0.1\,$dex. For the $\feh$-rich 
and -poor end, no gap has been detected, and we thus assumed a 
plateau (see \figurename{~\ref{thin_thick_transition}}). 
In the following, we identify stars with $\mgfe$ below 
that separation as \textup{\emph{thin}} disc stars and those
above that separation as \emph{thick} 
disc stars; we rejected stars with 
$\feh<-1.25\,$dex. The possible data contamination by halo stars is discussed 
at the end of Sect.~\ref{sub_re_sults}. The separation found by 
\citet{ges_disc_recio_2014} and \citet{Sarunas} with GES 
iDR1 data is thus again confirmed with the GES iDR2 data. We 
emphasize the fact that the two iDR1 thin- to thick-disc separations 
were found in the $\alffe-\mh$ and $[\text{Mg}/\text{M}]-\mh$ plane, 
respectively, where M is the global metallicity. As a consequence, our 
$\feh-\mgfe$ division is slightly different from these two previous 
studies, but totally compatible.

\section{Deriving the velocity dispersion of the Galactic disc}\label{deriv_veloc_disp}
This section describes in detail our estimation of the 
stellar velocity dispersion as a function of the chemical 
abundance $\mgfe$, which was adopted as a stellar clock.

For the \emph{main} sample, we decomposed the $\feh-\mgfe$ 
plane into 2D bins. First, the $\feh$ axis was cut from 
$\feh=-0.65$ to $\feh=+0.15\,$dex into four bins with a 
constant $0.2\,$dex step. The $\feh$-poor less populated 
tail was included in a single bin from $\feh=-1.25$ 
to $-0.65\,$dex, and we adopted the same procedure 
for the $\feh$-rich end from $\feh=+0.15$ to $+0.65\,$dex. 
In total, six $\feh$ bins were thus considered.
Second, we cut the $\mgfe$ axis into bins of $0.1\,$dex 
starting from $\mgfe=-0.2\,$dex up to $\mgfe=+0.4\,$dex, and 
we considered a last bin including all the stars with 
$\mgfe>+0.4\,$dex. We note that for each $\feh$ 
bin, the thin- to thick-disc transition presented in 
Sect.~\ref{stellar_sample} divides one $\mgfe$ bin into two 
distinct sub-samples.

For a given 2D bin, the average Galactocentric velocity $\mu$ and the 
intrinsic velocity dispersion $\sigma_i$ were derived by model fitting 
with a maximum likelihood approach. We computed the likelihood as follows:

\begin{equation}\label{eq_fwhm}
L(\mu, \sigma_i) = \prod_{i=1}^{N}\frac{1}{\sqrt{2\pi(\sigma_i^2 + ev_i^2)}}\,exp\left(-\frac{1}{2}\frac{(v_i - \mu)^2}{(\sigma_i^2 + ev_i^2)}\right), 
\end{equation}

where $v_i$ is the measured 
Galactocentric velocity for a star in a given velocity component and $ev_i$ its measured error. 
The measured velocity distribution can be seen as the convolution between the intrinsic distribution 
and the measured errors. In an equivalent manner, we minimised the log-likelihood 
$\Lambda \equiv -2\ln{L}$ with respect to both the mean velocity $\mu$ and the intrinsic velocity dispersion $\sigma_i$. 
To find the minimum, we investigated where the first-order derivative of $\Lambda$ goes to 
zero, which simultaneously solves the equations $\frac{\partial \Lambda}{\partial \mu}\mathbf{=0}$ and 
$\frac{\partial \Lambda}{\partial \sigma_i}\mathbf{=0}$, resulting in

\begin{equation}\label{eq_fwh}
\sum_{i=1}^{N}\frac{v_i}{(\sigma_i^2 + ev_i^2)} - \mu \sum_{i=1}^{N}\frac{1}{(\sigma_i^2 + ev_i^2)} = 0,
\end{equation}

and

\begin{equation}\label{eq_fwhm}
\sum_{i=1}^{N}\frac{(\sigma_i^2 + ev_i^2)^2 - (v_i - \mu)^2}{(\sigma_i^2 + ev_i^2)^2} = 0 
,\end{equation}

respectively (see also for instance \citealt{godwin_1987, pryor_1993}). 
This was performed using IDL procedures.

The quality of the kinematical data is crucial for the 
velocity dispersion estimation because the errors in the distances 
and proper motions propagate in the velocity 
measurements and sometimes create strong outliers. Therefore, we additionally cleaned our $\emph{main}$ sample so that it includes only stars with an accuracy in $\vr$, 
$\vf$ , and $\vz$ better than $30\,$\kms. We rejected $5\%$ of the stars that had extreme values of the velocity distribution, which were too far from the mean velocity. 
Finally, we estimated $\sigma_e$, the error on 
$\sigma_i$ , as the standard deviation of $1\,000$ bootstrap realisations. For a given bin $B$ 
containing $N$ stars, we created a new bin composed of $1\,000\times B$ stars. Then, we 
randomly selected $1\,000$ times $N$ stars in this new bin to measure $1\,000$ values of $\sigma_i$.

%ici1

\begin{figure}
\centering
\includegraphics[width=1.0\linewidth]{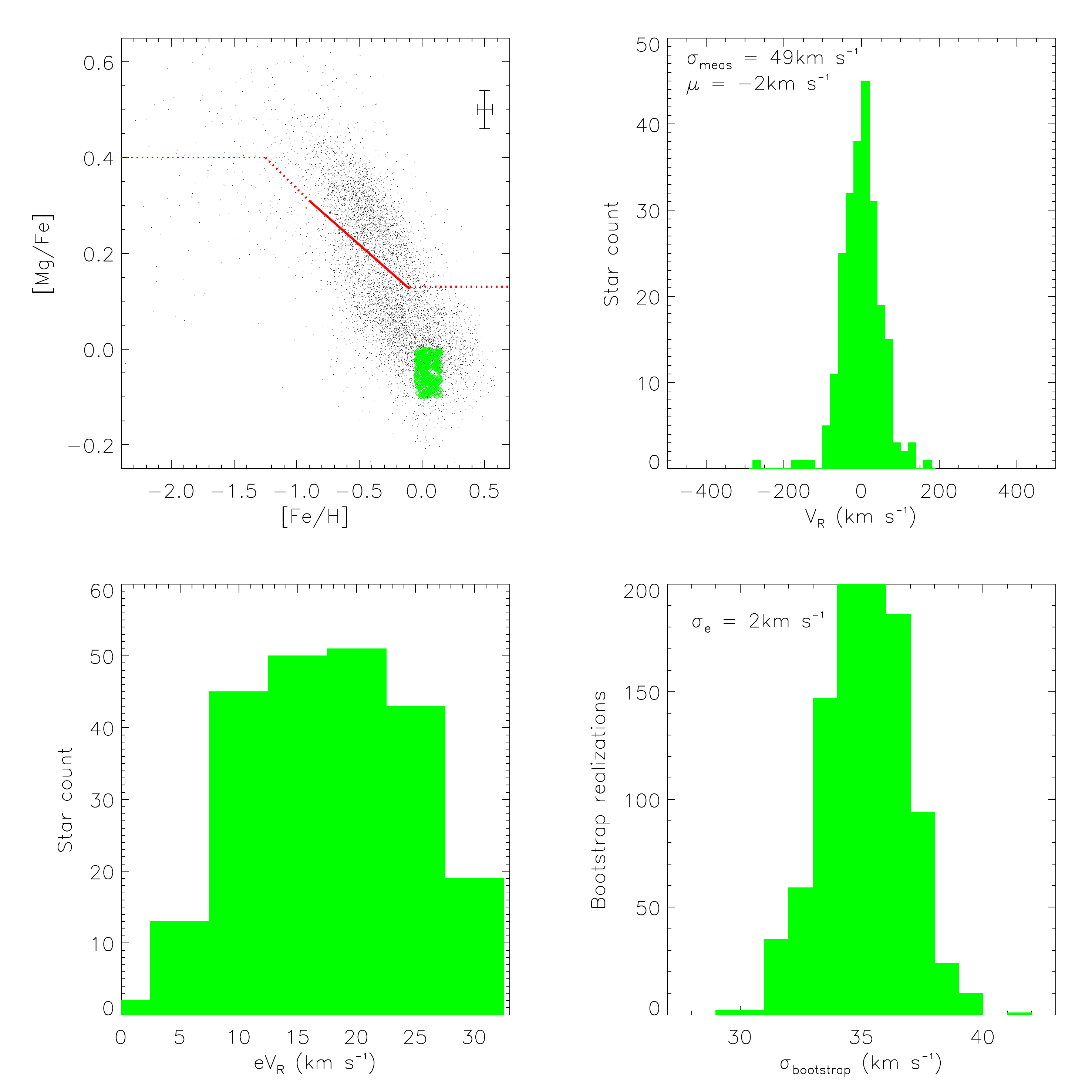}
\caption{\label{example_sigma}\emph{Top left:} $\mgfe$ as a function of 
$\feh$ for the \emph{main} sample. An example $\feh$-rich bin is shown in green (232 stars). 
\emph{Top-right:} Galactocentric radial velocity distribution for this bin with 
$\sigma_{\text{meas}}=49\,$\kms. 
\emph{Bottom left:} Distribution of errors on $\vr$.
\emph{Bottom right:} Velocity dispersion distribution corresponding to $1\,000$
bootstrap realisations, characterised by $\sigma_e=2\,$\kms. The 
intrinsic velocity dispersion derived by maximum likelihood is $\sigr=35\,$\kms.}
\end{figure}

As an illustration, we show in \figurename{~\ref{example_sigma}} the 
steps of the procedure we used to retrieve the velocity dispersion in the radial component 
for a given bin in the $\mgfe\,vs.\,\feh$ plane composed of 232 stars with $\langle\feh\rangle=+0.01\,$dex and 
$\langle\mgfe\rangle=-0.05\,$dex. Its velocity distribution is shown in the \emph{top right} panel. We find a measured 
velocity dispersion of $\sigma_{\text{meas}}=49\,$\kms. By model fitting with 
the maximum likelihood approach, we obtain $\sigr=35\,$\kms. 
A distribution of 1000 bootstrap realisations 
is shown in the \emph{bottom right} panel, characterised by a 
standard deviation of $\sigma_e=2\,$\kms.

\begin{figure}
\centering
\includegraphics[width=1.0\linewidth]{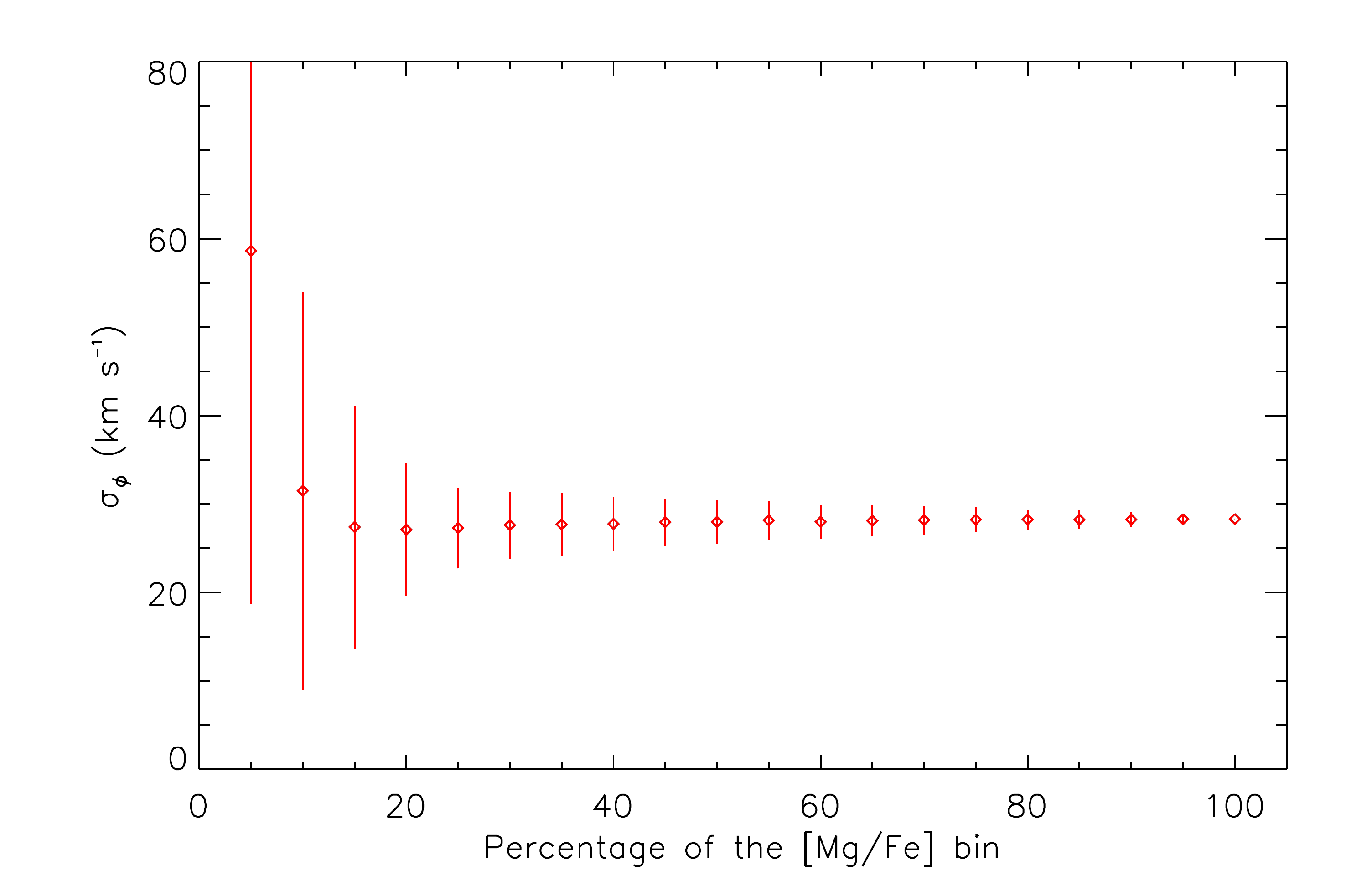}
\caption{\label{stabl} Average azimuthal velocity dispersion 
as a function of the selected percentage of a bin composed of 100 stars, 
characterised by $\langle\feh\rangle=-0.54\,$dex, $\langle\mgfe\rangle=+0.17\,$dex, and 
$\sigf=28\pm2\,$\kms. The error bars correspond to the 
standard deviation of 1000 bootstrap realisations.}
\end{figure}

We have tested the possible influence of the low 
statistics on the measured velocity dispersion. To this 
purpose, we selected a well-populated $\mgfe$ interval 
from the bin $\langle\feh\rangle=-0.54\,$dex (azimuthal 
component) composed of 100 stars ($\langle\mgfe\rangle=+0.17\,$dex, 
$\sigf=28\pm2\,$\kms). We randomly selected 1000 times 
a percentage of the stars, from $5$ to $100\%$ with a 
$5\%$ step. Then, we performed 1000 derivations of the 
velocity dispersion, computing an average and 
standard deviation. The resulting average of $\sigf$ is shown 
in \figurename~\ref{stabl} as a function of the percentage of 
the selected stars. We see that $\langle\sigf\rangle$ 
converges to a plateau after $10\%$ around the reference value 
$\sigf=28\,$\kms, with an increase of $30\,$\kms~for 
the lowest percentage (composed of only five stars). 
We also tested some other bins at different $\feh$ and 
$\mgfe$ ratios and observed the same behaviour. Based on this test, we conclude that bins with more than 15 stars are reasonable for a scientific application and 
are suitable for a reliable interpretation in terms of kinematical 
properties of the disc.

\begin{figure*}
\centering
\begin{tabular}[c]{c c c}
\includegraphics[width=0.3\linewidth]{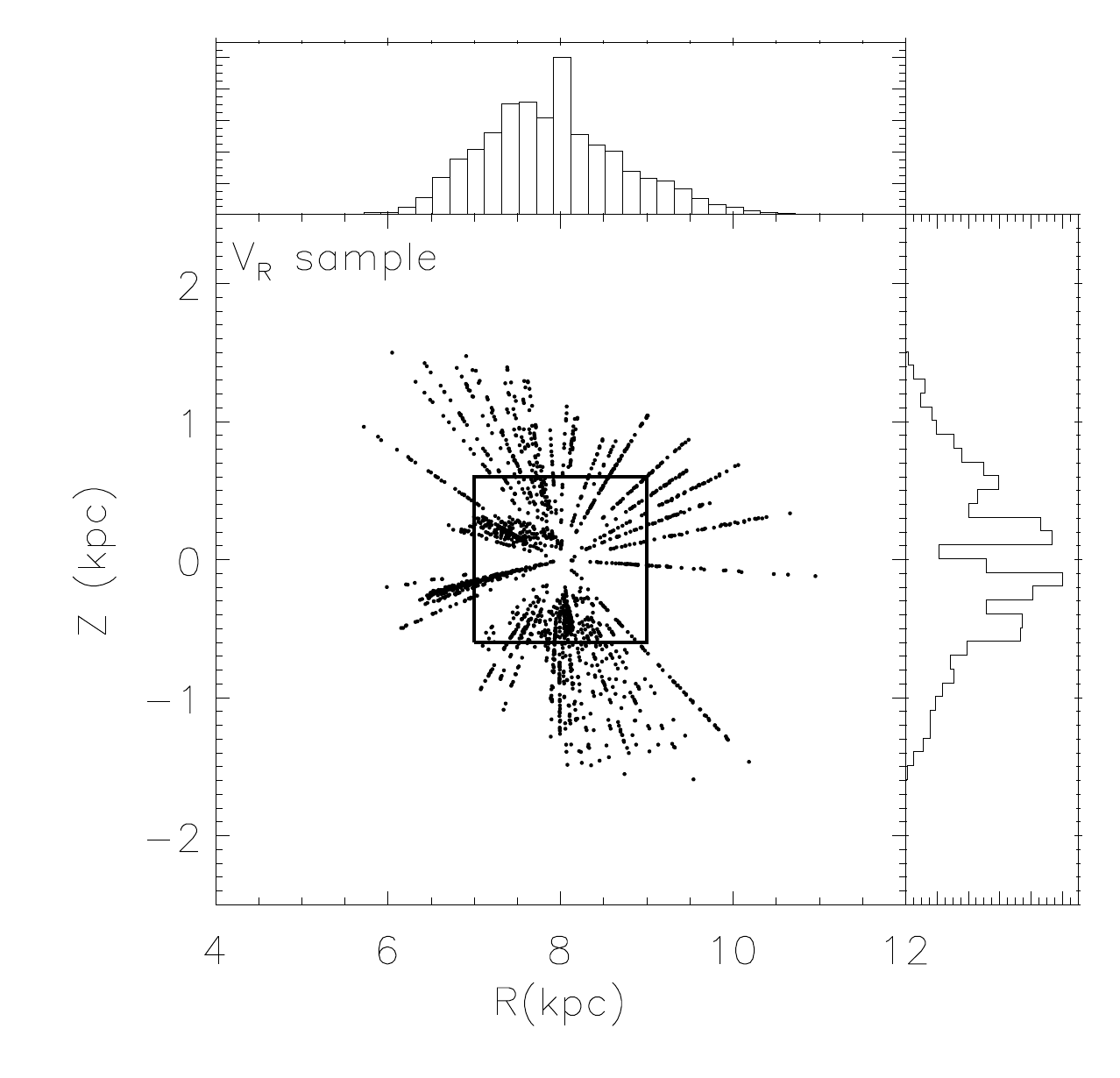} & \includegraphics[width=0.3\linewidth]{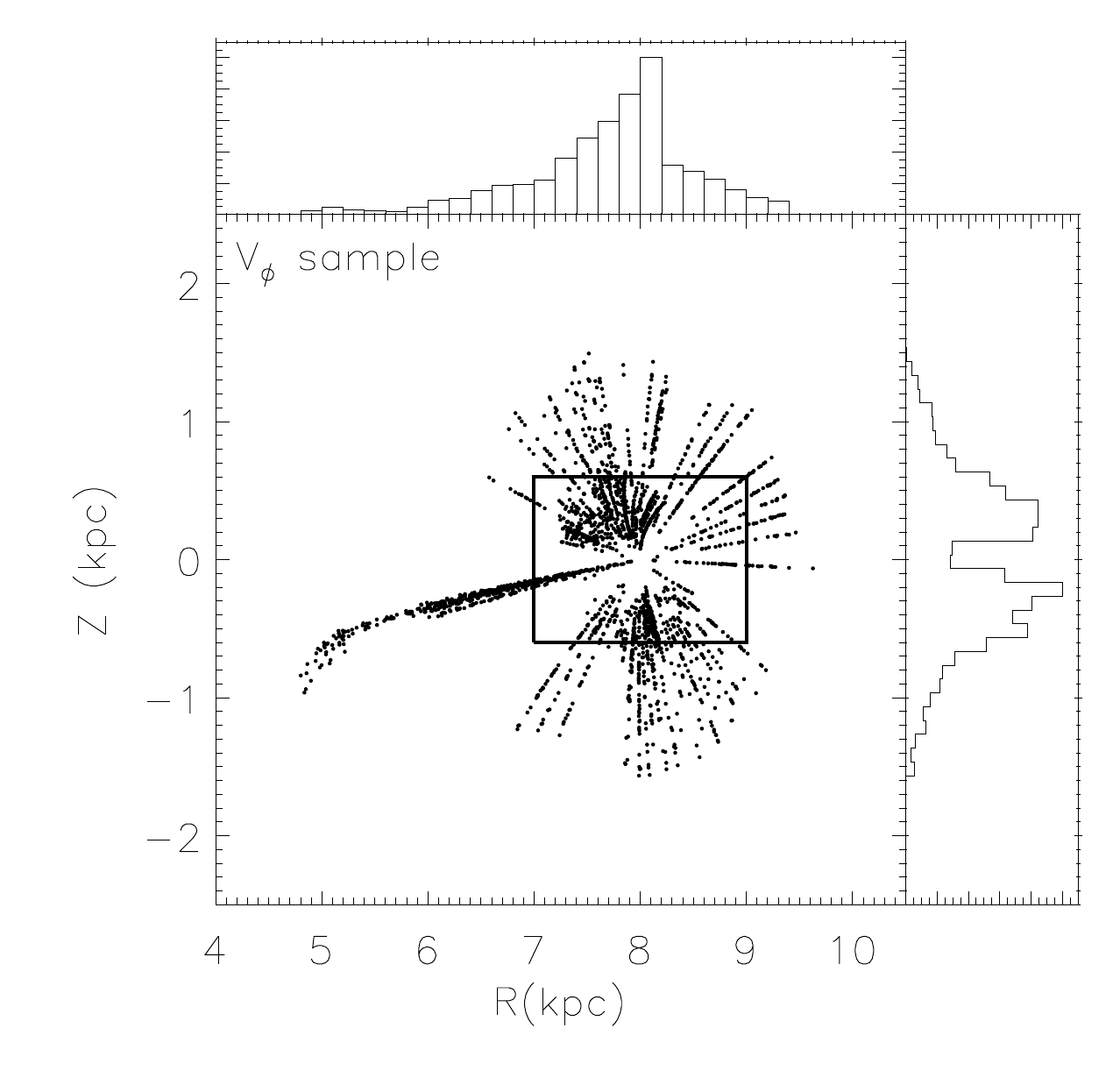} & \includegraphics[width=0.3\linewidth]{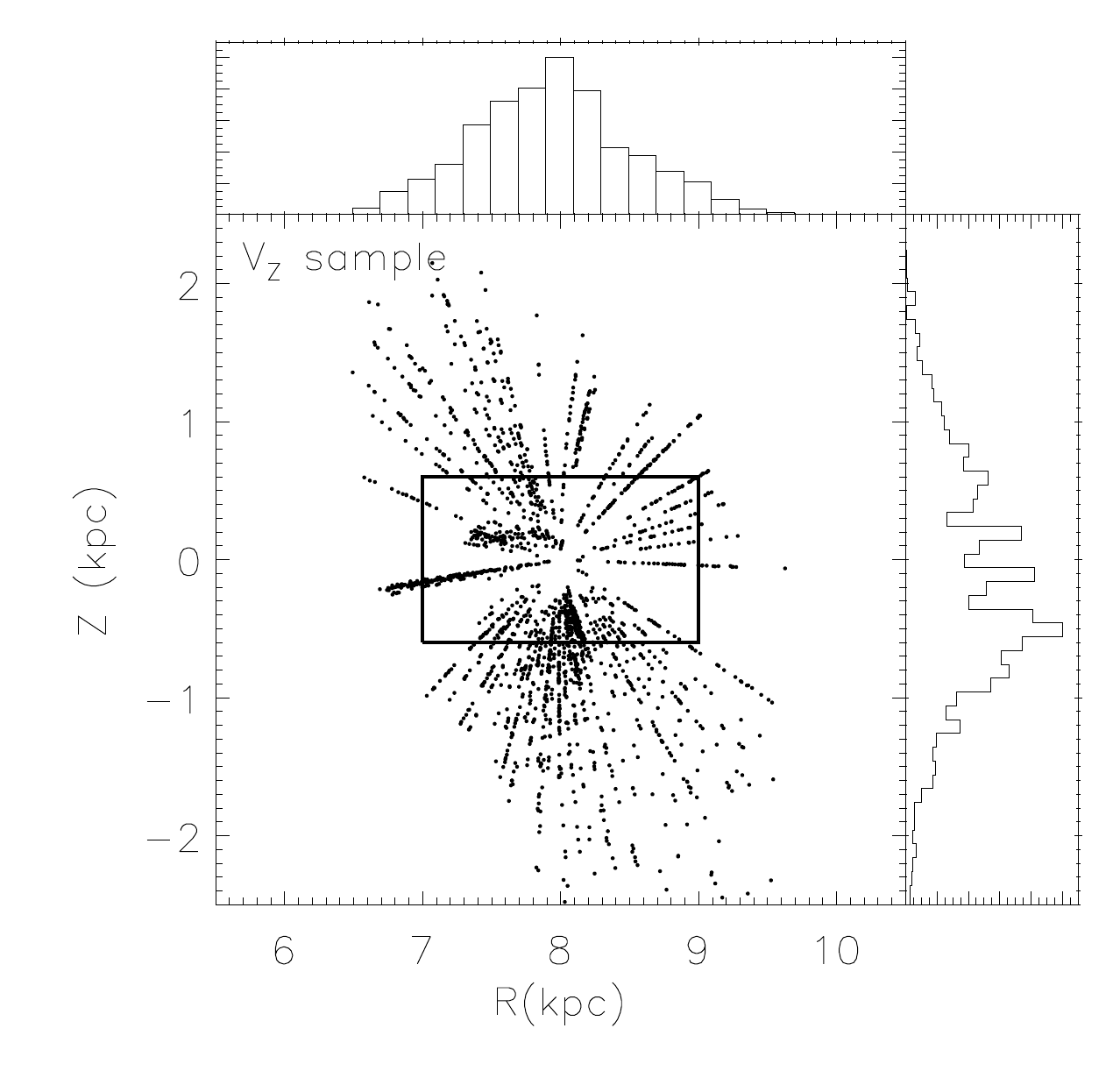} \\ 
\end{tabular}
\caption{\label{r_versus_z}Vertical distances from the Galactic plane ($Z$) as a function of the Galactic 
radius ($R$) for the radial, azimuthal, and vertical velocity components, composed of 
$1\,998$, $2\,459,$ and $2\,045$ stars, respectively. The black boxes correspond to the 
spatial coverage of the Solar suburb RAVE sample from \citet{minchev_chemo_2014}. We 
over-plotted the respective normalised distributions of R and Z.}
\end{figure*}

Finally, from our selection criteria in terms of Galactocentric velocities, 
we obtained three samples composed of $1\,998$, $2\,459,$ and $2\,045$ stars 
for the radial, azimuthal, and vertical Galactic velocity 
component, respectively, as shown in \figurename~\ref{r_versus_z}. 
These three samples quite homogeneously cover 
a wide Galactic radius range from 6 to $10\,\text{kpc}$, 
and $2\,$kpc above and below the Galactic plane. The fraction of stars 
belonging to the thin and thick discs as defined by the two sequences 
in elemental abundances is summarized in \tablename~\ref{table_thin_thick}. 
We see that the thin disc represents $\thicksim 80\%$ of the sample. We also 
verified that the mapping of the $\mgfe-\feh$ plane is 
the same for the three samples.

\begin{table}[h!]
\centering
\caption{Total number of stars ($N_{tot}$) for the three velocity components 
($R, \phi~\text{and}~Z$), with the respective proportion of the thin- and thick-disc 
samples.}
\begin{tabular}[c]{c c c c}
\hline
\hline 
Velocity component  &    $\vr$    &   $\vf$   &   $\vz$     \\
\hline 
$N_{tot}$  & $1\,998$  &  $2\,459$  &  $2\,045$ \\
Thin disc  & $1\,593$  &  $1\,982$  &  $1\,513$ \\
Thick disc &  $405$    &   $477$    &   $532$   \\
\hline
\end{tabular}
\label{table_thin_thick}
\end{table}

\section{Velocity gradients and velocity dispersions in the 
Galactic disc}\label{re_sults}

From the derived velocity dispersions, we first searched for correlations 
between the Galactocentric velocities with respect to $\feh$ 
and $\mgfe$ for the thin and the thick discs. We then derived the 
velocity dispersion relations with respect to $\mgfe$.

\subsection{Galactocentric velocities vs. $\feh$}\label{res_vel_fe}

\begin{figure*}
\centering
\includegraphics[width=1.0\linewidth]{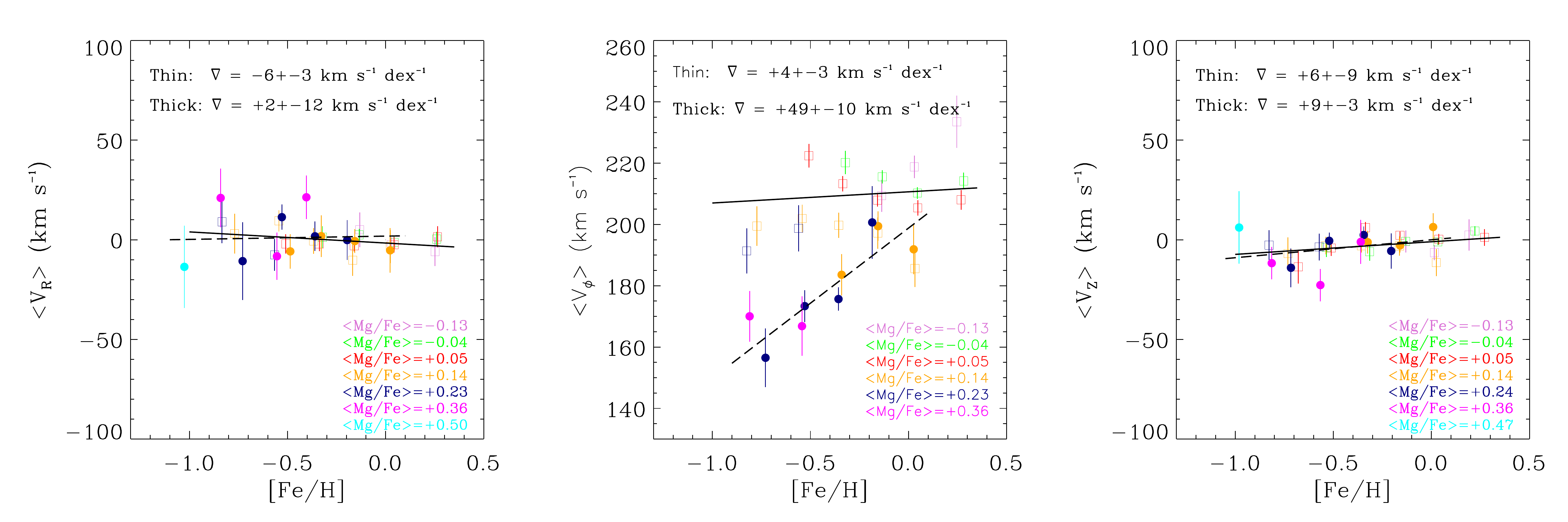}
\caption{\label{mean_v_feh} Average radial, azimuthal, and vertical 
velocity ($\meanvr$, $\meanvf,$ and $\meanvz$) as a function of the 
$\feh$ ratio. The $\mgfe$ bins are colour-coded, and the average 
values of each one are written in the legend. 
The thin and thick discs are represented by open squares ($\square$) 
and filled circles ($\bullet$), respectively. The error bars correspond 
to the standard errors on the mean velocity for each bin. The derived 
thin- and thick-disc gradients are shown as full and dashed lines, respectively.}
\end{figure*}

In \figurename~\ref{mean_v_feh}, we show the behaviour of the 
Galactocentric velocities as a function of the $\feh$ ratio. 
Each $\feh$ bin is also decomposed into several $\mgfe$ bins, 
following the same mapping of the $\mgfe-\feh$ plane as 
described in the previous section.

For the radial component (\emph{left} panel), we see that the 
thick disc shows no trend within the errors with $\feh,$ while the thin disc 
is characterised by a very weak negative gradient. We note that the 
scatter is larger for the most $\feh$-poor bins in which 
the number of stars is lower. The relation between the vertical 
Galactocentric velocity and the $\feh$ (\emph{right} panel) is similar 
in the thin and thick disc. Both thin and thick discs 
follow a unique sequence with a very weak slope, and in addition, 
the thin disc is compatible with no slope. Here again, the scatter 
of the lowest $\feh$ bins is due to the lower statistics. Finally, 
the behaviour of those two components agrees well with the 
expected trend for the Milky Way thin and thick discs.

In contrast, the azimuthal velocity (\emph{middle} panel) shows 
two clearly distinct sequences. The first one corresponds to the thin 
disc with a weak positive gradient equal to $+4\pm3\,$\kmsdex. The second 
sequence, belonging to the thick disc, is clearer with a 
stronger gradient $\gradfef=+49\pm10\,$\kmsdex. In both cases, 
the slopes and their associated errors were estimated through a linear 
regression procedure, taking into account the statistical errors of each 
bin. This azimuthal gradient estimated for the thick disc agrees 
with the one found by \citet{kordopatis_2011} ($\gradfef=45
\pm12\,$\kmsdex, even though their thick disc was not based on a chemical 
separation), and more recently, \citet{ges_disc_recio_2014}  
found $\gradfef=+43\pm13\,$\kmsdex\ with
GES iDR1 data. However, the thin-disc azimuthal 
gradient differs notably from previous studies. \citet{lee_2011b} and 
\citet{ges_disc_recio_2014} found a negative gradient of 
$-22.6\pm1.6\,$\kmsdex~and $-17\pm6\,$\kmsdex, respectively. In our 
study, the thin disc covers $-1.0<\feh<+0.5\,$dex, while in 
\citet{ges_disc_recio_2014}~and \citet{lee_2011b}, the range 
was smaller, $-0.9<\feh<+0.2$ and $-0.7<\feh<+0.3$, respectively. 
We first point out that the trend with $\feh$ masks a stronger 
one with $\mgfe,$ which is directly caused by the results being 
Fe-range dependent. Second, we discuss the robustness of 
the gradient with the Fe-range. Our lowest $\feh$ bin for the thin 
disc contains 59 stars ($\thicksim90\%$ of main sequence (MS) stars), 
while the highest $\feh$ bin contains 299 stars ($\thicksim80\%$ MS). 
We checked that those stars have a chemical pattern in silicon 
and calcium that is compatible with the one presented here in $\mgfe$. When we restrain 
our thin disc to a $\feh$ domain similar to that used by \citet{ges_disc_recio_2014} 
and \citet{lee_2011b}, \emph{i.e.} $-0.7<\feh<+0.2\,$dex, we find a 
negative gradient equal to $-5\pm5\,$\kmsdex. This result is consistent 
in terms of slope with the anti-correlations found previously. In a last check, 
we decided to compute a new gradient, for which we removed the stars in the thin-disc sequence 
with $\mgfe>0.15\,$dex. These stars are in the $\feh$-poor thin-
disc tail and show a different behaviour ($\vf$ values lower than 
$210\,$\kms, grey and blue curves) with respect to the other thin-disc members. We note that \citet{ges_disc_recio_2014} did
not take this 
population into account to compute the gradient because 
they lacked good statistics. In this case, we find a slope equal to 
$-8\pm4\,$\kmsdex, which is again consistent with the literature values. In 
summary, all these facts indicate that our azimuthal gradient is 
reliable. Therefore, we conclude that the thin-disc azimuthal 
gradient reflects the driving effect of $\mgfe$ because of
the dependence on the Fe-range.

\subsection{Galactocentric velocities vs. $\mgfe$}\label{res_vel_mg}

\begin{figure*}
\centering
\includegraphics[width=1.0\linewidth]{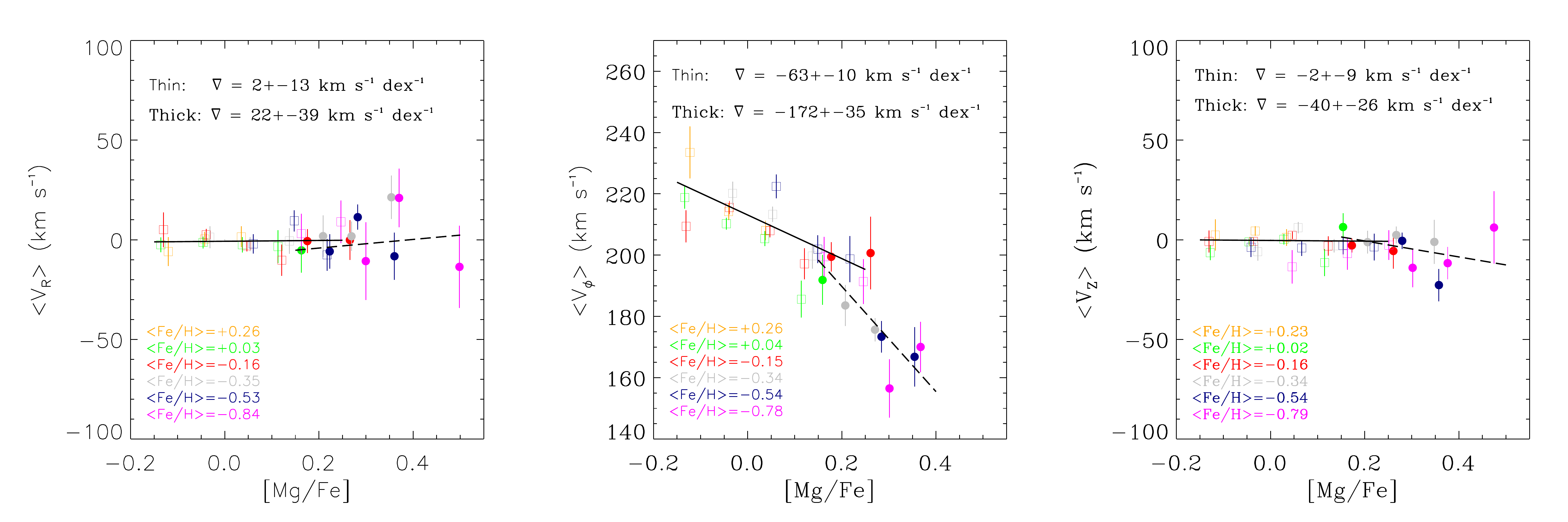}
\caption{\label{mean_v_mgfe}Average radial, azimuthal, and vertical 
velocity ($\meanvr$, $\meanvf,$ and $\meanvz$) as a function of the 
$\mgfe$ ratio. The $\feh$ curves are colour-coded, and the average values of each one 
are written in the legend. The symbols, errors bars, and linear fit are the same 
as in \figurename{~\ref{mean_v_feh}}.}
\end{figure*}

Figure~\ref{mean_v_mgfe} presents the Galactocentric velocities 
as a function of $\mgfe$. For the radial component (\emph{left} panel), we 
found no slope for the thin disc, while the thick-disc slope is null 
within the error. We note that the highest 
$\mgfe$ bins are more scattered as a result of lower statistics. In contrast, 
the vertical velocity (\emph{right} panel) shows a first sequence with no 
trend for the thin disc, while a second sequence, which corresponds to the thick 
disc, is characterised by a negative gradient equal to $-40\pm26\,$\kmsdex. 
Because we took into account the statistics errors to compute the 
slopes, this negative gradient is well explained by the presence of the 
lowest $\mgfe$ less populated bin. If we derive the slope without taking care 
of the errors on the mean velocity, we find a slope equal to $-21\pm30\,$\kmsdex. 
Thus, these gradients found for the radial and vertical average velocities 
with respect to $\mgfe$ are consistent with those obtained with respect to $\feh$.

The azimuthal average velocity (\emph{middle} panel) shows a strong 
anti-correlation with respect to $\mgfe$ that is in fact related to the 
increase of the velocity dispersion with age (see Sect.~\ref{impli_cations}). 
Moreover, the thin and thick discs show two distinct sequences. The thin 
disc is characterised by a negative gradient of $-63\pm10\,$\kmsdex~
(Spearman’s rank correlation coefficient equal to -0.73). Here again, 
we tried to quantify the impact of the $\feh$-poor and $\mgfe$-rich 
tail of the thin disc. When we removed this population, we found a gradient equal 
to $-61\pm11\,$\kmsdex, not significantly different from the full thin disc. 
The thick-disc anti-correlation is twice as steep and reaches 
$-172\pm35\,$\kmsdex~(Spearman’s rank correlation coefficient equal to -0.84). 
This azimuthal velocity gradient agrees very well with the one found 
by \citet{ges_disc_recio_2014} with GES iDR1 data ($\gradmgf=-181\pm38\,$\kmsdex).

\subsection{Galactic disc velocity dispersion vs. $\mgfe$}\label{sub_re_sults}
\begin{figure*}
\centering
\includegraphics[width=1.0\linewidth]{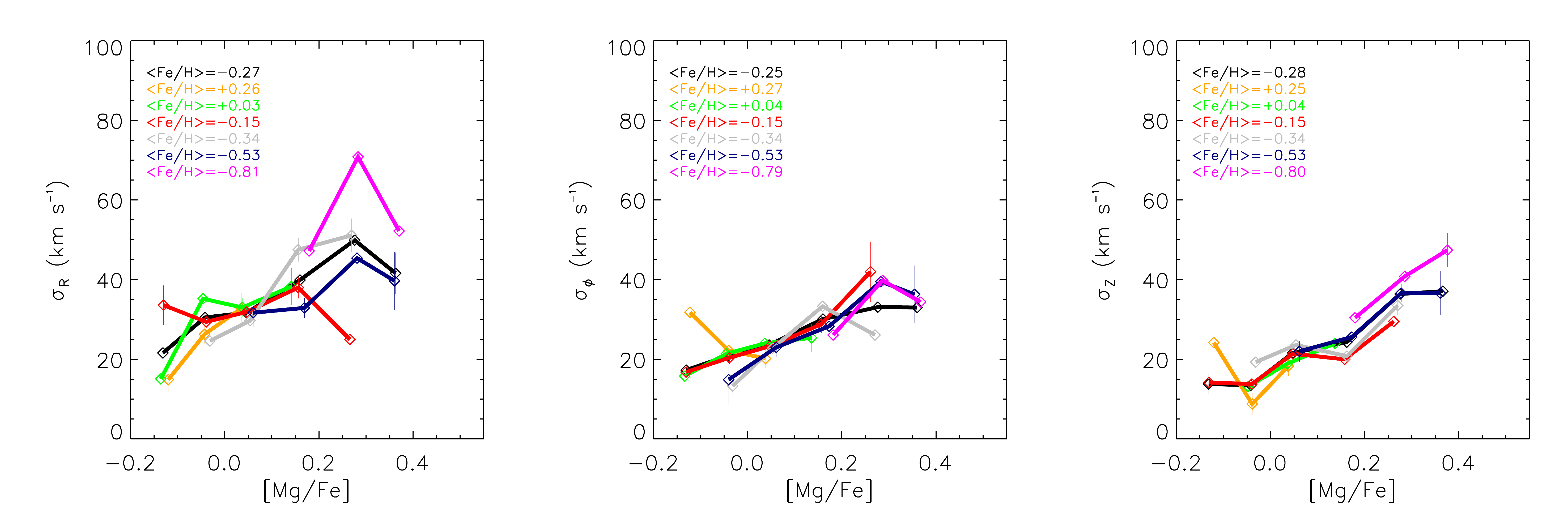}
\caption{\label{sigma_v}Radial, azimuthal, and vertical velocity 
dispersions ($\sigr$, $\sigf$ , and $\sigz$) as a function of 
$\mgfe$ ratio. The colour code and legend are the same as 
\figurename{~\ref{mean_v_mgfe}}. The errors bars come from the 
standard deviation of 1000 bootstrapping realisations (as explained in 
Sect.~\ref{deriv_veloc_disp}).}
\end{figure*}

We now first present the relation between the Galactic velocity 
dispersion and the $\mgfe$ ratio without considering the separation 
between the thin and thick discs to 
explore a possible continuity between the two components in terms 
of velocity dispersion. The resulting relation $\sigma\,vs\,\mgfe$ 
is shown in \figurename{~\ref{sigma_v}}. Again, the different 
$\feh$ curves are colour-coded, while the overall trend obtained 
considering all the $\feh$ range is represented in black.

For the radial velocity dispersion $\sigr$, two 
regimes are observed. First, $\sigr$ increases for all $\feh$ curves. 
The overall trend (in black) reaches from $21\pm3$ to $59\pm2\,$\kms~between 
$\mgfe=-0.13$ to $+0.28\,$dex. Second, for the most $\mgfe$-rich 
bins, the velocity dispersion seems to decrease. For the overall trend, 
a decrease of $9\,$\kms~is observed within $1-\sigma$. For the curve with 
$\feh=-0.15\,$dex (in red), $\sigr$ clearly decreases by $13\,$\kms~within 
$2-\sigma$. While for the curve with $\feh=-0.53\,$dex (in blue) the decrease is 
less obvious, $\sigr$ shows a strong drop for the most $\feh$-poor curve, 
$23\,$\kms~within $2-\sigma$. We note that all the $\mgfe$ bins showing 
a decrease are populated by more than 19 stars.

For the azimuthal velocity dispersion $\sigf$, a single 
behaviour seems to be observed. A smooth increase of the velocity 
dispersion is seen for the overall trend (in black) from $17\pm2$ to $33\pm3\,$\kms~throughout the $\mgfe$ range. The decreasing velocity dispersion of the highest 
$\mgfe$ bins observed for $\sigr$ is less evident here. A decrease
by $9\,$\kms~
is detected for the last $\mgfe$ sub-sample of the curve with $\feh=-0.34\,$dex, 
within $2-\sigma$. There are 104 stars. In addition, a drop of $\sigf$ is 
suggested for the two most $\feh$-rich curves (red and magenta), but it might equally well be 
interpreted as a plateau.

Finally, we measure a robust increase of $\sigz$ throughout 
the $\mgfe$ range. The overall trend (in black) reaches from $13\pm2$ 
to $37\pm3\,$\kms. The different curves of $\feh$ follow this main trend 
without showing any decrease as a function of $\mgfe$.

\begin{figure*}
\centering
\includegraphics[width=1.0\linewidth]{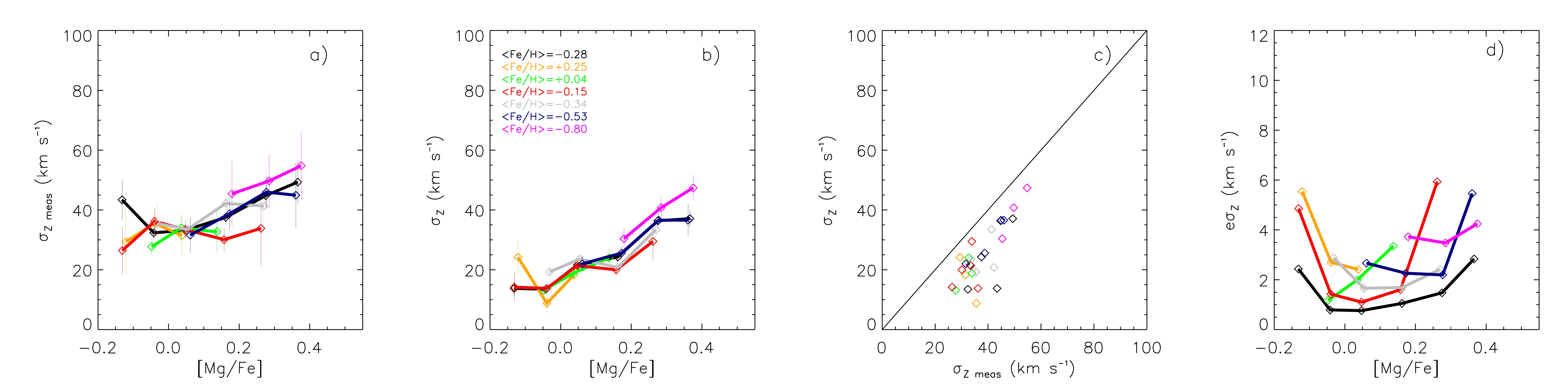}
\caption{\label{sigma_vz_quatre} $a)$ Measured vertical velocity dispersion $v.s.$ 
$\mgfe$. The error bars correspond to the standard error of the standard 
deviation. $b)$ Intrinsic vertical velocity dispersion $v.s.$ $\mgfe$, with 
error bars from the bootstrap realisations. $c)$ Measured $v.s.$ intrinsic vertical 
velocity dispersion. $d)$ Errors of the intrinsic vertical velocity dispersion 
$v.s.$ $\mgfe$.}
\end{figure*}

For the vertical velocity component, several quantities are plotted 
in \figurename{~\ref{sigma_vz_quatre}}. In panel a), the measured velocity dispersion 
is shown as a function of the $\mgfe$ ratio. For each bin, we computed a simple 
standard deviation of the sample, as presented in panel b). Both measured and intrinsic 
dispersions are compared in panel c), and as expected, $\sigma_{\text{meas}}$ is lower. 
The behaviour of the error on $\sigma_{\text{int}}$ is also shown in panel d). The errors are 
much smaller than the dispersion. For a given curve of $\feh$, the $U-$shape of the 
curve is due to the smaller statistics at each edge.

To check the reliability of all these velocity dispersion trends, we 
tested the effect of a slight shift by $\pm0.05\,$dex on the $\feh$ curves. For $\sigr$, 
both increasing and decreasing behaviours are recovered as a function of the $\mgfe$ ratio. 
For the azimuthal component, we still measure an increase of $\sigf$ , and the dispersion 
seems to reach a plateau for the highest bins in $\mgfe$. Finally, the smooth rise of 
$\sigz$ is recovered on average.

In addition, we checked the properties of the stars composing the bins that  show a 
decrease ($\mgfe>0.2\,$dex). On one hand, the median errors in $\tef$ and therefore in distances are higher 
than the typical thin-disc errors, but still lower than $55\,$K and $0.13\,$kpc, 
respectively. On the other hand, the median internal errors in $\logg$, $\feh$, 
$R,$ and Galactocentric velocities are totally comparable with the rest of the 
sample, and no peculiar lines-of-sight were chosen to observe these stars. 
We also checked the median internal error for each $\mgfe$ bin, which is around $0.04\,$dex, 
while the most Fe-rich/Mg-poor bin reaches $0.07\,$dex. Moreover, the standard error 
of the mean $\mgfe$ value for each bin is small, typically $0.006\,$dex, reaching 
$0.03\,$dex for the most Fe-poor curve. As a consequence, the velocity dispersion trends 
are not sensitive to the errors on $\mgfe$. Finally, as developed in 
Sect.~\ref{deriv_veloc_disp}, the velocity measurements for low-statistics samples 
do not reflect any bias leading to an artificial under-estimation 
because a minimal number of 15 stars per bin was fixed.\\

\begin{figure*}
\centering
\includegraphics[width=1.0\linewidth]{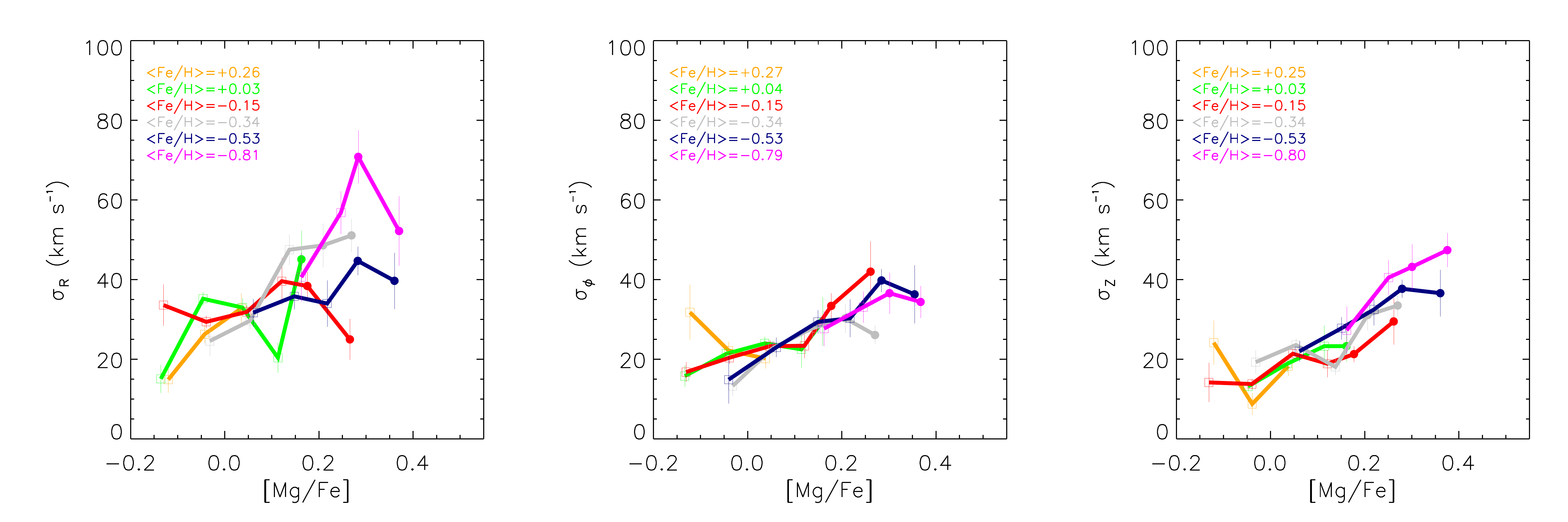}
\caption{\label{sigma_v_separ}Same as 
\figurename{~\ref{sigma_v},} but adopting the thin- to thick-disc separation as determined in Sect.~\ref{stellar_sample} 
(open squares and filled circles, respectively).}
\end{figure*}

Considering the thin- to thick-disc separation 
determined in Sect.~\ref{stellar_sample}, we computed the new velocity 
dispersions for each velocity component of each disc to see whether the chemical 
separation can highlight a possible dynamical distinction. The results are shown in 
\figurename~\ref{sigma_v_separ}. Compared to the previous plots 
without separation, we clearly see that regardless of the velocity component, 
the results are very similar. We note that for a given $\feh$ curve, a $\mgfe$ bin located 
at the thin- to thick-disc transition is divided into two smaller bins, with two values of the velocity 
dispersion on average equal to the entire bin (albeit with larger errors). As a result, we can 
determine that the thin- and thick-disc populations show very different behaviours, at least for 
$\sigr$. The thin-disc sequence follows the main increasing trend of $\sigr$, while the highest values 
of $\sigma$ and the decreasing bins of $\mgfe$ belong to the thick-disc 
stars. For $\sigf$ and $\sigz$ the thick discs can be seen as an extension 
of the thin disc. The velocity increases smoothly from the thin- 
to the thick-disc sequences without a major visible break.

As the volume covered by our data extends to distances quite far from the 
Galactic plane ($|Z|>1.5\,$kpc), we tested a possible contamination of our 
sample by halo stars that could have artificially increased the mean velocity dispersion 
of some metal-poor bins. To do so, we computed the velocity dispersion for 
three ranges of $Z$. First, we selected the stars with $|Z|<0.5\,$kpc 
($80\%$ of the sample). The typical increase of $\sigma$ is observed 
for all the velocity components, while the decreasing behaviour 
is not seen because of insufficient statistics in the higher $\mgfe$ bins.
Then, we selected the stars with $|Z|<1.0\,$kpc ($94\%$ of the sample), 
recovering most of the same trends as shown in \figurename~\ref{sigma_v} 
thanks to the increasing number of high-$\mgfe$ stars. Finally, computing 
the velocity dispersion for the stars with $|Z|<1.5\,$kpc (more than $99\%$ of 
the sample), we find the trends obtained \figurename~\ref{sigma_v},
as expected.
Therefore, the contamination from halo stars does not seem to be responsible for the observed 
patterns that seem to be present close to the plane, although with lower thick-disc star statistics. In addition, to bias the mean velocity dispersion estimations, 
the contamination would need to be in-homogeneous in $\mgfe$ for a given $\feh$ curve (more 
halo stars contaminating the intermediate $\mgfe$ bins than the high $\mgfe$ 
bins presenting lower dispersion values), which seems unlikely. Finally, we emphasize that the selection function is not optimised to target very metal-poor stars 
in large amounts, as is visible from the $\feh$ coverage of 
\figurename{~\ref{thin_thick_transition}} and \figurename{~\ref{example_sigma}}. 
We therefore conclude that the contamination by halo stars is not significant in our study.

In this section, we thus derived robust trends between 
the Galactocentric velocities with respect to $\feh$ and $\mgfe$ and
found a good consistency with previous studies (Sects.~\ref{res_vel_fe} and~\ref{res_vel_mg}). Moreover, 
we established a robust relation between $\mgfe$ and the 
Galactocentric velocity dispersion (Sect.~\ref{sub_re_sults}). This relation is derived here for the 
first time for a very broad spatial coverage, probing both 
above and below the Galactic plane and including the Solar 
neighbourhood. We also showed that the stars with decreasing 
velocity dispersion belong entirely to the thick-disc sequence.

\section{Effect of the spatial coverage}\label{dis_cussion}

In this section, we first compare the $\mgfe-\sigma$ 
relation established above with the recent results by 
\citet{minchev_chemo_2014} for the RAVE survey. Then we study 
in a second part the impact of the $R-Z$ 
coverage on our $\mgfe-\sigma$ relation.

\subsection{Comparison with the RAVE survey}
In a recent paper, \citet{minchev_chemo_2014} studied the velocity 
dispersion of the Galactic disc based on RAVE DR4 data 
\citep{kordopatis_2013}. We recall that the resolution of the RAVE 
survey is lower than that of GES: $\text{R}\thicksim7000$ against 
$\text{R}\thicksim18000$. Furthermore, RAVE established the relation 
between the velocity dispersion and the $\mgfe$ ratio for giants ($\logg<3.5$) 
with a typical uncertainty of $0.15\,$dex in $\mgfe$ \citep{boeche_2011}, 
larger than the ones adopted in the present work based on higher resolution spectra. 
Minchev and collaborators focused on the Solar suburb, selecting stars in the range 
$|Z|<0.6\,$kpc and $7<R<9\,$kpc. We emphasize that 
our study covers a larger Galactic volume with half the sample size 
of the study of Minchev et al. Our results obtained with GES iDR2 
data (\figurename{~\ref{sigma_v}}) are quite compatible with the work of 
\citet{minchev_chemo_2014}. On one hand, we recover on average the same smooth increase of 
$\sigma_{R, \phi, Z}$ as a function of $\mgfe$, regardless of the velocity component for 
$\mgfe<+0.1\,$dex, followed by a steeper increase of $\sigma_{R, \phi, Z}$ up to 
$\mgfe<+0.3\,$dex. We also detect a decreasing behaviour for the most $\feh$-poor 
and $\mgfe$-rich stars for the radial component. However, we are able to measure a 
clear decrease of $\sigr$ for the curve with $\feh=-0.15\,$dex because of the higher 
precision in $\mgfe$. On the other hand, contrary to \citet{minchev_chemo_2014}, we do not detect any decrease of $\sigz$ 
with $\mgfe$, finding a smooth continuous increasing trend. We note that the lack of GES stars with $\mgfe>+0.4\,$dex 
may be responsible for not resolving the decrease observed by \citet{minchev_chemo_2014} 
in this range of $\mgfe$.

\begin{figure*}
\centering
\includegraphics[width=1.0\linewidth]{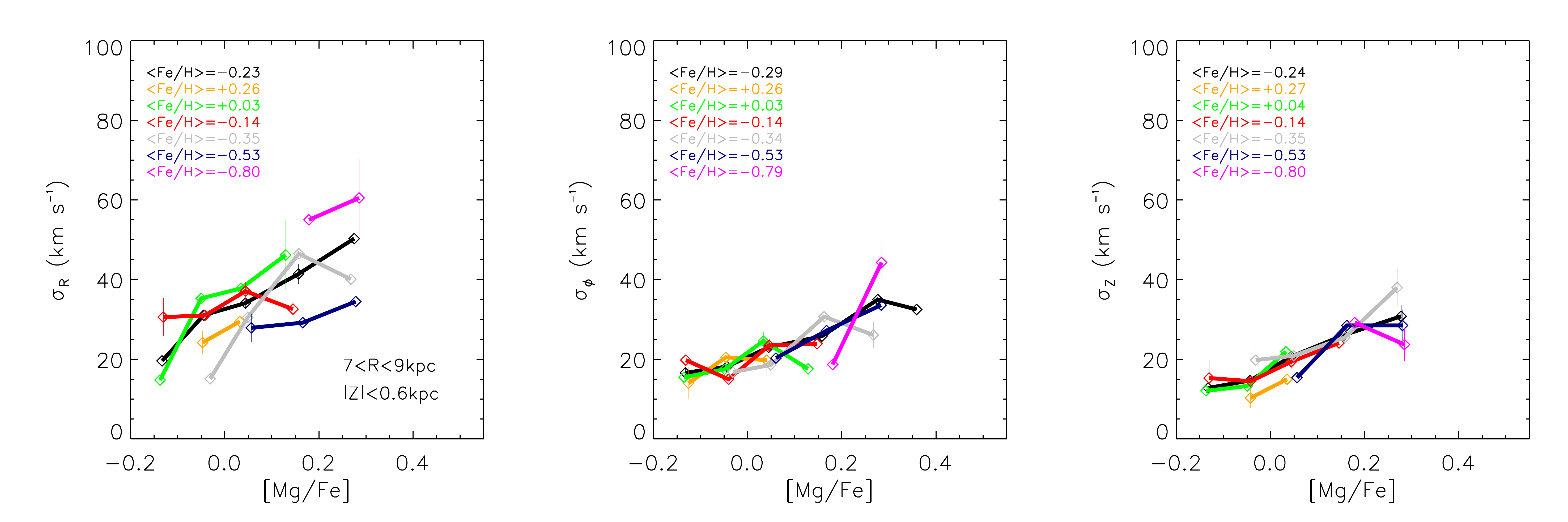}
\caption{\label{sigma_v_separ_rave}Same as 
\figurename{~\ref{sigma_v},} but the spatial coverage corresponds to the one 
adopted by \citet{minchev_chemo_2014} for the Solar suburb. We note that 
in this case, the number of stars in our samples is a factor of 4-5 smaller than 
the one used by \citet{minchev_chemo_2014}, which results in not resolving the 
decline in $\sigz$.}
\end{figure*}

To complete our comparison, we now derive a new $\sigma - \mgfe$ relation, 
but selecting stars with the same $\text{R}-\text{Z}$ coverage as in 
\citet{minchev_chemo_2014} (see the boxes in \figurename{~\ref{r_versus_z}}). We 
kept exactly the same mapping of the $\mgfe-\feh$ plane, as well as 
the same conditions on the Galactocentric velocity errors. For each 
velocity component, the number of stars was then reduced, corresponding 
to 48, 54, and 44 $\%$ of the three samples presented 
at the end of Sect.~\ref{deriv_veloc_disp}. We show our results in 
\figurename{~\ref{sigma_v_separ_rave}}. First, we see that the stellar 
populations with low $\feh$ and high $\mgfe$ tend to disappear. There is no statistics to derive a robust velocity dispersion value, 
the bins sometimes contain no more than three to four stars. Second, the overall trend of $\sigma$ increases smoothly with 
$\mgfe \   $ for the three velocity components, as expected. Moreover, $\sigr$ seems to decrease for the highest $\mgfe$ bin 
of the grey curve with $\feh=-0.35\,$dex. Despite the lower number of bins 
available to analyse with this sub-sample of the data, we see that the 
properties of the velocity dispersion observed with RAVE are generally compatible 
with the iDR2 GES data over the same Galactic volume. The GES data allow us 
to extend these behaviours to a much larger volume.

\subsection{Probing the $R-Z$ space}\label{probe_rz}
The broad spatial coverage of the GES iDR2 data allows analysing the possible 
dependencies of the velocity dispersion on the position of the stars in the 
Galaxy ($R$ and $Z$).

\begin{figure*}
\centering
\includegraphics[width=1.0\linewidth]{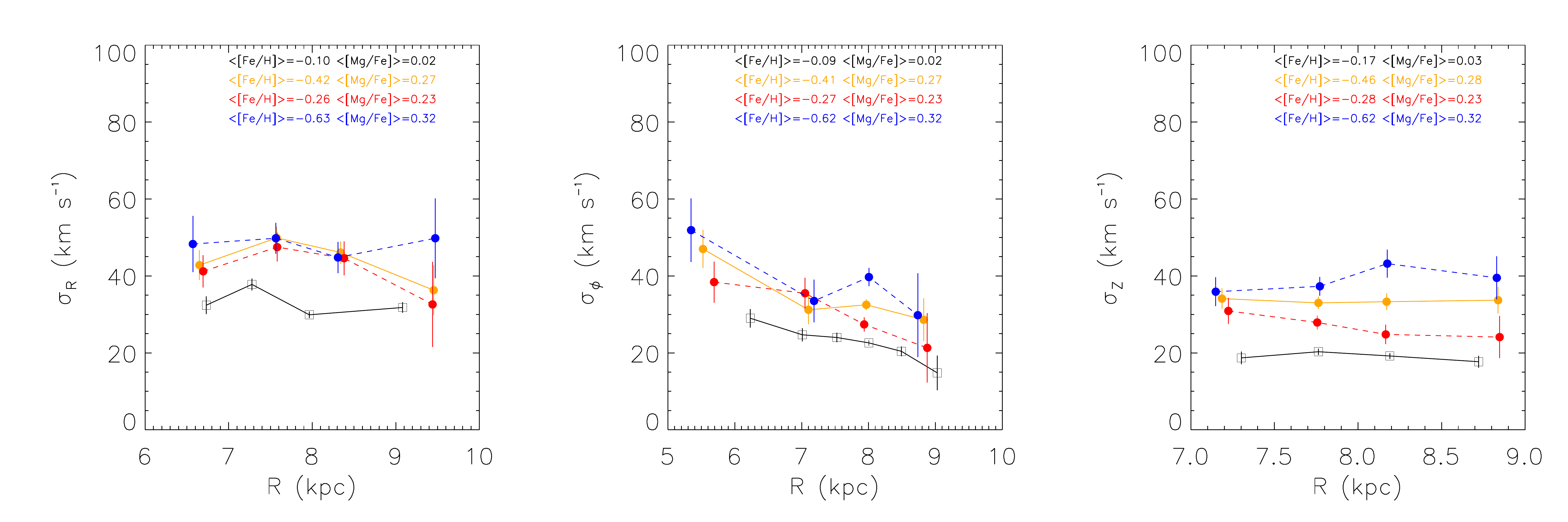}
\caption{\label{sigma_v_vs_R}Radial, azimuthal, and vertical velocity dispersion 
($\sigr$, $\sigf$ , and $\sigz$) as a function of the Galactocentric radius R. The 
thin and the thick discs are represented by open squares and filled 
circles, respectively. The full orange line corresponds to the whole 
thick disc, while the red and blue dashed lines represent its decomposition into 
two sub-samples (Fe-rich and Fe-poor). The error bars correspond to the 
standard deviation of 1000 bootstrapping realisations. The average $\feh$ and 
$\mgfe$ values of each population for the whole $R$ range are marked.}
\end{figure*}

For this purpose, we first computed the velocity dispersion of each Galactic 
velocity component as a function of the Galactic radius $R$, taking into 
account the chemical thin- to thick-disc separation (see Sect.~\ref{stellar_sample}).
 In addition, following the results of Sect.~\ref{sub_re_sults}, we divided the 
thick disc into two sub-samples for $\feh\leq-0.45\,$dex and 
$\feh>-0.45\,$dex. We show our results in \figurename{~\ref{sigma_v_vs_R}}. 
First, the behaviour of $\sigr$ with $R$ for the thin and thick discs 
shows no clear trend; it is nearly constant around 35 and $45\,$\kms, respectively. 
For the thick disc, we observe that the velocity dispersion of the $\feh$-poor population 
is higher than the $\feh$-rich population. Second, $\sigf$ shows 
a clear decrease with R for both discs. Within $2-\sigma$, the thin disc is characterised by  a 
decline of $\sim15\,$\kms~over the whole range of $R$, while the thick disc seems to decrease 
by $\sim18\,$\kms~in the same range. For $\sigz$, no particular trend seems to emerge for either the thin 
or the thick disc. We conclude that the most significant trends belong to the thin and the thick disc 
in the azimuthal component with a decrease of $\sigf$ with increasing Galactocentric 
radius. We also note that the trend in $\sigf$ and $\sigz$ may be 
partially hidden because the samples in $\vf$ and $\vz$ do not extend as far in $R$ 
as the one in $\vr$ ($\vf$ extends more towards the centre and lacks the anti-centre, while 
$\vz$ is restricted to $7-9\,$kpc). However, a clear decrease is seen for the thick-disc 
azimuthal velocity with increasing radius (see \figurename{~\ref{mean_v_vs_R_vs_Z}}.). 
We also find a rough estimate of $6.5\,$kpc for the radial scale length 
of the azimuthal velocity dispersion in the whole thick disc. This estimate agrees with \citet{lewis_1989} ($R_{\sigma_{\phi}}^{\text{thick}}=6.7\,$kpc).
We note that $\sigz$ is expected to decrease with $R$ (see for example the study 
of \citet{bovy_2012_vertical} with the SEGUE data); we do not
observe this behaviour here. We checked the 
case where we considered the disc as a whole (no thin- to thick-disc separation) and measured a constant trend of 
$\sigz$ with R. From the trends \figurename{~\ref{sigma_v_vs_R}}, we measured the ratio 
$\sigr/\sigz$ for both the thin and the thick disc at $R=8\,$kpc, finding $\sigr/\sigz\sim0.7$ for both discs. 
Our sample being composed of cool dwarfs and giants, this value agrees well with $\sigr/\sigz\sim0.6$ 
of \citet{binney_2014} with the RAVE survey, considering no thin- to thick-disc separation.

\begin{figure*}
\centering
\includegraphics[width=1.0\linewidth]{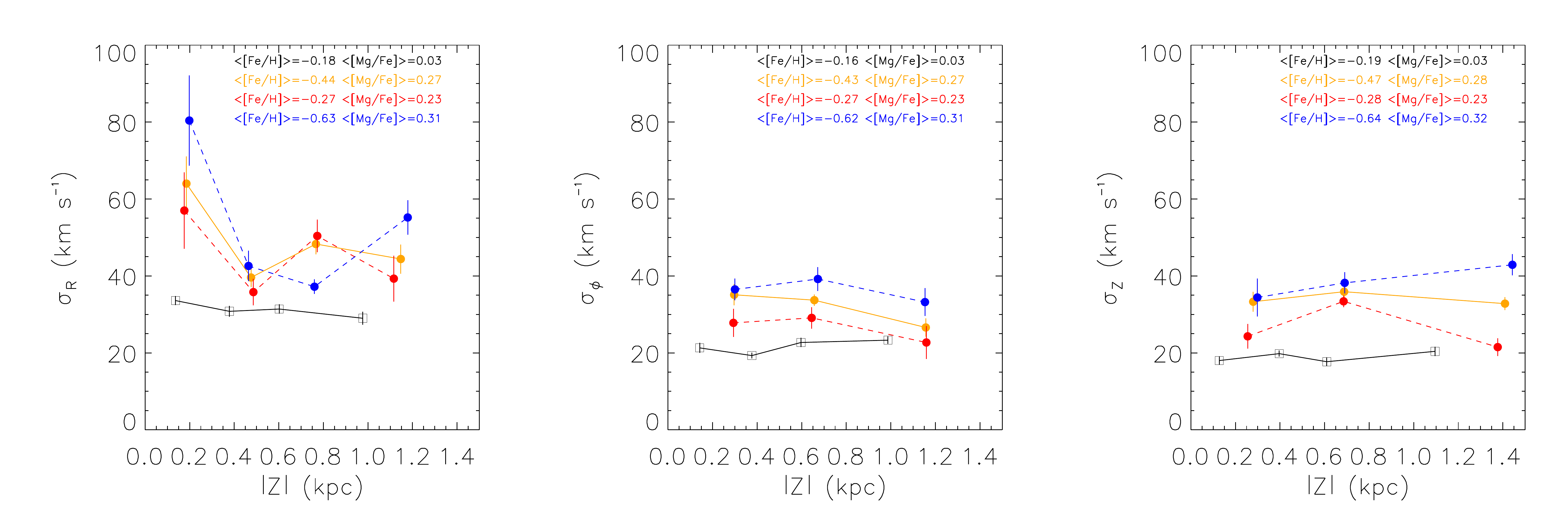}
\caption{\label{sigma_v_vs_Z}Radial, azimuthal and vertical velocity dispersion 
($\sigr$, $\sigf$ and $\sigz$) as a function of the height to the Galactic plane. 
The symbols, colours, and error bars are the same as in \figurename{~\ref{sigma_v_vs_R}}.}
\end{figure*}

\begin{figure}
\centering
\includegraphics[width=0.8\linewidth]{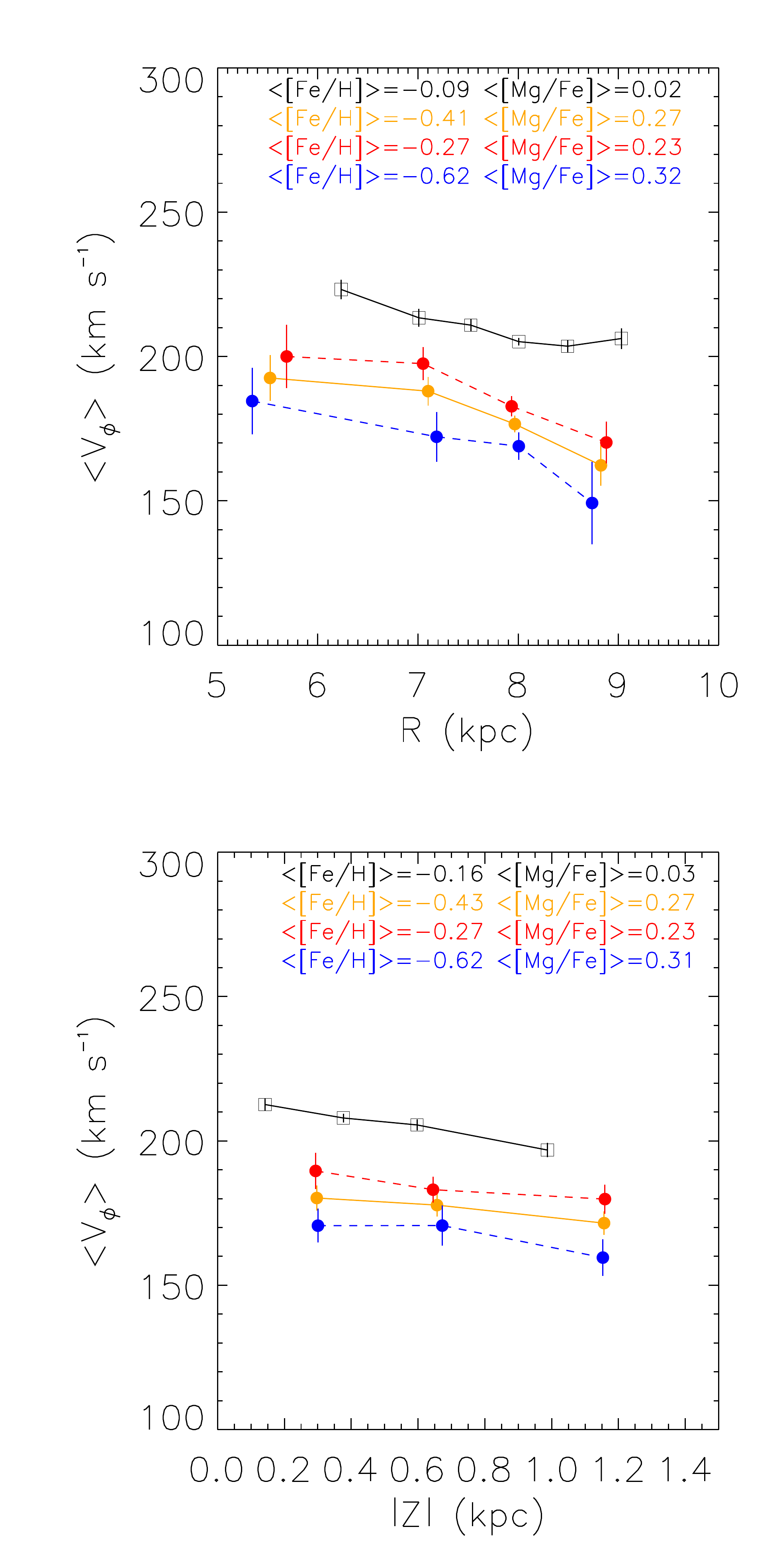}
\caption{\label{mean_v_vs_R_vs_Z}\emph{Top:} Mean azimuthal velocity 
$<\vf>$ as a function of the Galactic radius. \emph{Bottom:} $<\vf>$ as a 
function of the height to the Galactic plane. The symbols and colours are
 the same as in \figurename{~\ref{sigma_v_vs_R}}. The error bars correspond 
to the standard error on the mean velocity.}
\end{figure}

We then determined the behaviour of the velocity dispersion with the 
distance to the Galactic plane. To avoid possible observational 
biases, for example the radial trend of $\sigma$ presented in the previous 
paragraph, we selected the stars in the range $7<R<9\,$kpc. The results are shown in 
\figurename{~\ref{sigma_v_vs_Z}}. The thick-disc stars are divided into the same 
two sub-samples as in \figurename{~\ref{sigma_v_vs_R}}. First, from a global 
point of view, we did not detect any significant trend with $|Z|$ for the thin disc 
(black curve). While $\sigr$ seems to decrease with $Z$, $\sigf$ and $\sigz$ tend to show a higher 
velocity dispersion at larger $Z$. Second, we found that the velocity dispersion of the thick disc 
is higher than that of the thin disc, regardless of the distance to the Galactic plane. The complete 
thick-disc sample (orange curve) does not show any trend for $\sigr$. In addition, $\sigz$ is 
quite constant (isothermal) for the thin disc, which could be the signature of stars with 
nearly circular orbits. A decrease of $10\,$\kms~within $2-\sigma$~is observed for 
$\sigf$ in the range of $Z$ covered. On the other hand, the trends followed 
by the two thick-disc sub-samples are more scattered in the radial component. We 
note the high-velocity dispersion of the thick-disc bins at $|Z|<0.3\,$kpc. 
For $\sigf$ and $\sigz$, the two thick-disc populations are very well separated 
with no particular trends. We thus conclude that the most significant trend is 
the possible decrease of $\sigf$ with $Z$ for the thick disc, while the thin disc does not 
seem to follow any trend, apart from the higher values of the velocity dispersion at large $Z$. 
We checked the corresponding values of the thick-disc azimuthal velocity 
for the studied bins and found a weak gradient $\gradvphiz=-10\pm7\,$\kmskpc~within 
the $1-\sigma$ error (see \figurename{~\ref{mean_v_vs_R_vs_Z}}). \citet{kordopatis_2011} found 
a negative gradient $\gradvphiz=-19\pm8\,$\kmskpc, but the range in $Z$ was four times 
broader. A significant decrease of $\vf$ with $Z$ was also observed by \citet{ges_disc_recio_2014} 
with GES iDR1 data for a different and less restricted selection of stars, only for $|Z|>1.5\,$kpc. 
\citet{binney_2014} observed a strong increase of $\sigr$ and $\sigz$ with $Z$ in the RAVE data, and 
we do not observe such a strong increase in our data, probably because we separated the disc into two 
thin- and thick-disc components. As a test, we checked the behaviour of $\sigma_{R, \phi, Z}$ with $Z$ without 
separating it into thin- and thick-disc parts. On one hand, we observe a constant behaviour of $\sigr$ with $Z$ around $34\,$\kms. 
On the other hand, $\sigf$ and $\sigz$ increase over the whole range of $Z$ of $6$ and $11\,$\kms, respectively. The 
increasing trend of $\sigz$ with Z therefore agrees with the observations of \citet{binney_2014} with 
the RAVE data and also with the chemo-dynamical model of \citet{mcm_2013}.

\section{Implications on the Galactic disc formation}\label{impli_cations}

In this section, our results are discussed in the context of 
the different Galactic disc formation scenarios. First of all, our 
analysis can be summarized in the following points:

\begin{enumerate}
\item[-] We chemically identified two distinct populations in the $\mgfe\,vs.\,\feh$ 
plane and defined them as the thin and thick disc (see Sect.~\ref{stellar_sample}).

\item[-] As detailed in Sects.~\ref{res_vel_fe} and~\ref{res_vel_mg}, we confirm 
that there is no correlation between $\vr$ and $\vz$ with respect to $\feh$ or to $\mgfe$ for either 
the thin or thick disc. However, we found a weak positive gradient of $\vf$ with 
$\feh$ for the thin disc, while the thick disc shows a strong positive trend. 
Moreover, $\vf$ is strongly anti-correlated with $\mgfe$ for both populations.

\item[-] The thin-disc velocity dispersion increases with $\mgfe$ and seems to 
display  a single regime (see Sect.~\ref{sub_re_sults}).

\item[-] The thick-disc velocity dispersion seems to show a single decreasing trend 
with increasing $\mgfe$ at least for the radial component (see 
Sect.~\ref{sub_re_sults}). The decline is observed for three curves in $\feh$ 
for $\sigr$ (between 10 and $20\,$\kms), while it is only suggested for $\sigf$. 
No decline is observed in $\sigz$.

\item[-] For the thick disc, $\sigf$ decreases as a function of $R$ by 
$18\,$\kms, while $\sigf$ seems to decline ($10\,$\kms) with $Z$ 
(see Sect.~\ref{probe_rz}).
\end{enumerate}

The thin- and thick-disc dichotomy is still a matter of debate 
\citep{bovy_2012_no_thick}. On one hand, the similarity 
in the physical properties between the $\feh$-rich thick disc and the thin disc around 
$\feh\thicksim-0.15\,$dex suggests a common evolution scenario at a certain epoch 
in the history of the two components. This behaviour was also observed by 
\citet{ges_disc_recio_2014} and \citet{Adibekyan2011} from the behaviour of the $\alffe$ 
abundances. Moreover, a recent study by \citet{nidever_2014_apogee} with the APOGEE 
survey separated the Galactic disc into an $\alpha$-rich and an $\alpha$-poor population 
and highlighted an overlap between these populations at $\feh\thicksim+0.2\,$dex. In 
our study, this common behaviour is reinforced by the similar azimuthal and vertical 
velocity dispersions in the metal-rich regime between the two sequences. On 
the other hand, the chemical distinction of the two populations in the 
$\mgfe\,vs.\,\feh$ plane for $\feh<-0.20\,$dex suggests two different evolutionary paths 
in the formation of the Galactic disc. Moreover, we clearly detected two regimes in our 
chemo-kinematical analysis, both in the study of the Galactocentric velocity trends with 
$\feh$ and $\mgfe$. The $\vf$ anti-correlation of the $\mgfe$-low part of the thin disc with $\feh$ may be considered 
as a consequence of an outward migration of these stars due to epicyclic motions with their birth radii 
\citep{loebman_2011}. The $\mgfe$-rich tail azimuthal velocity of the thin disc shows no trend with $\vf$, 
suggesting a different evolutionary path than for the rest of the thin disc. 
In addition, both the thin and the thick discs show an anti-correlation of $\vf$ with 
$\mgfe$, with a break at $\mgfe=0.2\,$dex. As the velocity dispersion is expected to 
increase with age, older stars are expected to show lower $\vf$ (see for example Fig. 16 from 
\citet{Haywood_2013}). Here, the $\mgfe$ plays 
the role of the age, and both gradient are due to the driven effect of the age-velocity 
dispersion relation. The knee is due to the break in the age$-\mgfe$ relation (as shown 
by \citet{Haywood_2013}).

In terms of velocity dispersion, while the thin disc seems to have slightly 
increasing trend with $\mgfe$, the thick disc shows a higher dispersion with a 
clear decrease for the higher $\mgfe$ values at different $\feh$ values. \citet{minchev_chemo_2014} 
first detected this unexpected decrease of the velocity dispersion 
with RAVE data. They proposed that several mergers of decreasing intensity heated the outer disc 
(which would correspond to the highest values of velocity dispersion). These mergers led to subsequent radial 
migration of stars from the inner disc, resulting from the induced spiral structure. Stars from 
the inner disc migrating outwards would show the detected cooler kinematics (see Fig. 2 in 
\citealt{minchev_2014b}). We note that this scenario is valid only if the migration did not 
have accompanying kinematic heating and that stars on more circular orbits preferentially migrate.
In agreement with this scenario, the stars from our study that
show this 
decline do not seem to be preferentially located in the inner ($R<8\,$kpc) or in the outer 
($R>8\,$kpc) parts of the Galactic disc, but are homogeneously distributed as the rest 
of the sample in terms of line of sight. 
Moreover, the low $\sigr$ observed for the sub-sample with $\langle\feh\rangle=-0.15\,$dex 
and $\mgfe=+0.27\,$dex is significant in our data, corresponding to the metal-rich thick disc. According to Fig. 4 from \citet{minchev_chemo_2014}, the age of these stars with $\mgfe\approx0.27\,$dex 
would be around $6\,$Gyr with a birth radius smaller than their actual positions (according to their 
simulation). The larger metallicity coverage of the decrease presented in our data was nevertheless 
already predicted by the simulations of \citet{minchev_chemo_2014} (see their Fig. 3). We point 
out that the decrease of $\sigz$ at $\mgfe>+0.4\,$dex in their study is not observed in our GES data.

We note that the method of mapping the 2D abundance plane developed in both 
our study and in that of Minchev et al. is similar to the approach of "mono-abundance populations" 
by \citet{bovy_2012_vertical}. They found that $\sigz$ increases for SEGUE dwarf stars with 
high $\mgfe$ and low $\feh$, but no decrease for these stars is observed. Moreover, \citet{liu_2012} 
measured an increase of $\sigz$ with $\feh$ and $\mgfe,$ but the relation is less clear for $\mgfe-$rich 
stars.

On the other hand, the fact that the older stars ($\feh<-0.15\,$dex and 
$\mgfe>+0.20\,$dex) show a low velocity dispersion, comparable to that of 
the youngest thin-disc stars (\figurename{~\ref{sigma_v_separ}}, \emph{left} panel), 
indicates that these objects were born in a cool environment, that is to say\emph{}, were shielded 
from strong merger perturbations in the inner disc, where the density is high, and 
were not heated over the subsequent $10\,$Gyr. Moreover, we observe the velocity dispersion 
decline for different $\feh$ bins, reflecting different ages and stellar generations. 
This fact traces the chemical enrichment in iron of the disc. 

We can also compare our study with the recent work of \citet{Haywood_2013}, who 
investigated the age structure of the stellar populations in the close Solar neighbourhood. 
Their scenario is based on a thick-disc formation through a turbulent phase marked by an intense star 
formation. No decrease of the velocity dispersion was observed by these authors for the 
more $\feh$-poor stars (the oldest) of the thick disc, although we stress that their statistics 
is much smaller than that of our study. One of their main conclusions was that the radial 
migration, in the sense of churning (e.g. \citealt{schonrich_2009}) did not play a fundamental 
role in the thick-disc formation. \citet{Haywood_2013} also argued that the thick disc remained confined to the 
inner disc. The peak velocity dispersion that we observe for 
$\mgfe\thicksim0.3\,$dex is compatible with an initial turbulent disc with higher 
random velocities. Nevertheless, this picture has to somehow leave room for the existence 
of a small proportion of thick disc stars with cool kinematics (those in the high-$\mgfe$ 
bins with the observed dispersion decrease).

On the other hand, the cool kinematics of the Mg-rich tail of the thick-disc stars does not support 
the idea of accreted satellites, as proposed by \citet{abadi_2003}. The velocity dispersion 
of the older stars in our study is too low compared to the expected behaviour of the thick-disc 
stars in their simulation ($>80\,$\kms). The presence of a cool thick-disc tail could not be 
interpreted by the formation mechanism proposed by \citet{bournaud_2009}
either, who evoked a clumpy star 
formation. The proposed perturbations in the disc lead to a scatter of both the stars and the 
gas, increasing the velocity dispersions without room for the existence of stars with cool kinematics 
in the higher $\mgfe$ bins at each $\feh$ ratio. The scenario suggested by \cite{brook_2004}, 
who argued that a gas-rich merger was responsible of the thick-disc formation, cannot explain the 
presence of such dynamically cool stars either. In their numerical simulation, the thick-disc 
phase is characterised by a strong increase of the stellar velocity dispersion with age, again 
in contradiction with the trends observed here.

Finally, we can discuss our results in the context of the scenario proposed by 
\citet{villalobos_2008}, who argued that the thick disc was built 
by a pre-existing disc that was heated by a minor merger. Their model predicts a 
remnant-disc with cool kinematics, composed of old stars, with a typical location below 
$|Z|=1\,$kpc. These old stars would roughly correspond in our study to the bins with 
$\feh<-0.5\,$dex and $\mgfe>0.3\,$dex. For a given $\feh$ bin, these stars are characterised 
by cooler kinematics, and $~80\%$ of the stars are located below $|Z|=1\,$kpc. These stars 
might be good candidates to represent this remnant-primary old disc, because it is 
not affected by the merger 
event. However, the fact that the stars with cooler kinematics show a particular chemical 
pattern (higher $\mgfe$ values at each $\feh$ bin) than the stars with 
hot kinematics presents a difficulty for this scenario. On the other hand, the possibility 
of a chemical segregation between the heated and the non-heated stars could be considered. 
In particular, the old confined disc might have been more efficient in forming stars and recycling metals 
than the heated disc. 

\section{Conclusion}\label{con_clusion}
From the internal data release 2 of the Gaia-ESO Survey, we selected a sample 
of $7\,800$ FGK stars to study the chemo-kinematical structure of 
the Galactic disc. We derived the kinematics of these 
stars based on the fundamental parameters and proper motions. Based on the $\feh$ and $\mgfe$ 
abundances, we found a chemical gap between a first population with $\langle\feh\rangle=-0.15\,$dex and 
$\langle\mgfe\rangle=+0.04\,$dex and a second group with $\langle\feh\rangle=-0.45\,$dex and 
$\langle\mgfe\rangle=+0.29\,$dex that we associated with the thin- and thick-disc sequences, respectively.

We derived the trends between the three-dimensional Galactic velocities and the 
$\mgfe$ and $\feh$ ratios. We found a weak correlation between $\vf$ and $\feh$ 
for the thin disc due to the presence of a $\feh$-poor tail, while a strong correlation 
was established for the thick disc, in agreement with previous studies.

Located within $6<R<10\,$kpc and $|Z|<2\,$kpc, the GES iDR2 data allowed us to derive a 
relation between the Galactic velocity dispersion and the $\mgfe$ ratio, covering therefore 
a large Galactic disc volume. In terms of velocity dispersion, on one hand, 
the thin disc shows a smooth increase of $\sigma$ with increasing $\mgfe$. On the other hand, 
the thick disc can be seen as a continuity of the thin disc and reflects a decrease of its 
radial velocity dispersion for the stars with the highest values of $\mgfe$. This is the second 
time that this behaviour is observed; it was first discovered by \citet{minchev_chemo_2014} with RAVE 
and SEGUE data. The low dispersion observed for the most $\mgfe$-rich bins suggests 
that these stars were formed in a relatively quiet environment. Moreover, our observations 
could be explained by the scenario of \citet{minchev_chemo_2014}:\textit{} a series of merger 
events perturbing and heating the outer disc, inducing spiral arms that propel stars outwards 
in the disc. On the other hand, other proposed scenarios of thick-disc formation, including 
external or internal mechanisms, need to be refined to take into account this new observational 
constraint: the existence of a sub-sample of old disc stars with cool kinematics that show a 
particular and distinct chemical pattern, but are embedded in a generally turbulent context.

Finally, we conclude that combining individual chemical abundances and kinematics is a 
powerful tool for studying the stellar populations of the Milky Way, provided that both the 
chemical and kinematical estimates are derived with high enough precision. On the 
other hand, we worked in the context of the $\mgfe$ used as an age proxy, and we note that we cannot exclude that if the $\mgfe$-age correlation fails, the stars presenting cool kinematics 
and high-$\mgfe$ values could be formed at recent epochs, which
would completely change the picture of the 
interpretation scenarios. Further investigations are necessary to probe an always larger volume 
with an increasing number of stars, for example with the next data releases of the Gaia-ESO Survey and 
the spatial mission Gaia from the European Space Agency. The possibility of deriving precise 
stellar ages to directly study the age-velocity dispersion will be a huge qualitative 
improvement in our comprehension of the Galactic disc evolution.

\begin{acknowledgements}
We are grateful to the referee for the constructive comments that have helped us 
to improve our paper. G. Guiglion and A. Recio-Blanco thank Paola di Matteo and Misha 
Haywood for their careful reading of the paper and constructive comments. Based on data 
products from observations made with ESO Telescopes at the La Silla Paranal Observatory 
under programme ID 188.B-3002. These data products have been processed by the Cambridge 
Astronomy Survey Unit (CASU) at the Institute of Astronomy, University of Cambridge, and 
by the FLAMES/UVES reduction team at INAF/Osservatorio Astrofisico di Arcetri. These data 
have been obtained from the Gaia-ESO Survey Data Archive, prepared and hosted by the Wide 
Field Astronomy Unit, Institute for Astronomy, University of Edinburgh, which is funded by 
the UK Science and Technology Facilities Council. This work was partly supported by the 
European Union FP7 programme through ERC grant number 320360 and by the Leverhulme Trust 
through grant RPG-2012-541. We acknowledge the support from INAF and Ministero dell' 
Istruzione, dell' Universit\`a' e della Ricerca (MIUR) in the form of the grant "Premiale 
VLT 2012". The results presented here benefit from discussions held during the Gaia-ESO 
workshops and conferences supported by the ESF (European Science Foundation) through the 
GREAT Research Network Programme. A. Recio-Blanco, P. de Laverny and V. Hill acknowledge 
the ”Programme National de Cosmologie et Galaxies” (PNCG) of CNRS/INSU, France, for financial 
support. 
\end{acknowledgements}

\bibliographystyle{aa}
\bibliography{disc_cite}

\begin{thebibliography}{57}
\expandafter\ifx\csname natexlab\endcsname\relax\def\natexlab#1{#1}\fi

\bibitem[{{Abadi} {et~al.}(2003){Abadi}, {Navarro}, {Steinmetz}, \&
  {Eke}}]{abadi_2003}
{Abadi}, M.~G., {Navarro}, J.~F., {Steinmetz}, M., \& {Eke}, V.~R. 2003, \apj,
  597, 21

\bibitem[{{Adibekyan} {et~al.}(2011){Adibekyan}, {Santos}, {Sousa}, \&
  {Israelian}}]{Adibekyan2011}
{Adibekyan}, V.~Z., {Santos}, N.~C., {Sousa}, S.~G., \& {Israelian}, G. 2011,
  \aap, 535, L11

\bibitem[{{Allende Prieto} {et~al.}(2008){Allende Prieto}, {Majewski},
  {Schiavon}, {Cunha}, {Frinchaboy}, {Holtzman}, {Johnston}, {Shetrone},
  {Skrutskie}, {Smith}, \& {Wilson}}]{allende_2008}
{Allende Prieto}, C., {Majewski}, S.~R., {Schiavon}, R., {et~al.} 2008,
  Astronomische Nachrichten, 329, 1018

\bibitem[{{Bensby} {et~al.}(2003){Bensby}, {Feltzing}, \&
  {Lundstr{\"o}m}}]{bensby_2003}
{Bensby}, T., {Feltzing}, S., \& {Lundstr{\"o}m}, I. 2003, \aap, 410, 527

\bibitem[{{Bensby} {et~al.}(2014){Bensby}, {Feltzing}, \& {Oey}}]{bensby_2014}
{Bensby}, T., {Feltzing}, S., \& {Oey}, M.~S. 2014, \aap, 562, A71

\bibitem[{{Bergemann} {et~al.}(2014){Bergemann}, {Ruchti}, {Serenelli},
  {Feltzing}, {Alves-Brito}, {Asplund}, {Bensby}, {Gruyters}, {Heiter},
  {Hourihane}, {Korn}, {Lind}, {Marino}, {Jofre}, {Nordlander}, {Ryde},
  {Worley}, {Gilmore}, {Randich}, {Ferguson}, {Jeffries}, {Micela},
  {Negueruela}, {Prusti}, {Rix}, {Vallenari}, {Alfaro}, {Allende Prieto},
  {Bragaglia}, {Koposov}, {Lanzafame}, {Pancino}, {Recio-Blanco}, {Smiljanic},
  {Walton}, {Costado}, {Franciosini}, {Hill}, {Lardo}, {de Laverny}, {Magrini},
  {Maiorca}, {Masseron}, {Morbidelli}, {Sacco}, {Kordopatis}, \& {Tautvai{\v
  s}ien{\.e}}}]{bergemann_2014}
{Bergemann}, M., {Ruchti}, G.~R., {Serenelli}, A., {et~al.} 2014, \aap, 565,
  A89

\bibitem[{{Binney} {et~al.}(2014){Binney}, {Burnett}, {Kordopatis},
  {Steinmetz}, {Gilmore}, {Bienayme}, {Bland-Hawthorn}, {Famaey}, {Grebel},
  {Helmi}, {Navarro}, {Parker}, {Reid}, {Seabroke}, {Siebert}, {Watson},
  {Williams}, {Wyse}, \& {Zwitter}}]{binney_2014}
{Binney}, J., {Burnett}, B., {Kordopatis}, G., {et~al.} 2014, \mnras, 439, 1231

\bibitem[{{Boeche} {et~al.}(2011){Boeche}, {Siebert}, {Williams}, {de Jong},
  {Steinmetz}, {Fulbright}, {Ruchti}, {Bienaym{\'e}}, {Bland-Hawthorn},
  {Campbell}, {Freeman}, {Gibson}, {Gilmore}, {Grebel}, {Helmi}, {Munari},
  {Navarro}, {Parker}, {Reid}, {Seabroke}, {Siviero}, {Watson}, {Wyse}, \&
  {Zwitter}}]{boeche_2011}
{Boeche}, C., {Siebert}, A., {Williams}, M., {et~al.} 2011, \aj, 142, 193

\bibitem[{{Bournaud} {et~al.}(2009){Bournaud}, {Elmegreen}, \&
  {Martig}}]{bournaud_2009}
{Bournaud}, F., {Elmegreen}, B.~G., \& {Martig}, M. 2009, \apjl, 707, L1

\bibitem[{{Bovy} {et~al.}(2012{\natexlab{a}}){Bovy}, {Rix}, \&
  {Hogg}}]{bovy_2012_no_thick}
{Bovy}, J., {Rix}, H.-W., \& {Hogg}, D.~W. 2012{\natexlab{a}}, \apj, 751, 131

\bibitem[{{Bovy} {et~al.}(2012{\natexlab{b}}){Bovy}, {Rix}, {Hogg}, {Beers},
  {Lee}, \& {Zhang}}]{bovy_2012_vertical}
{Bovy}, J., {Rix}, H.-W., {Hogg}, D.~W., {et~al.} 2012{\natexlab{b}}, \apj,
  755, 115

\bibitem[{{Bovy} {et~al.}(2012{\natexlab{c}}){Bovy}, {Rix}, {Liu}, {Hogg},
  {Beers}, \& {Lee}}]{bovy_2012_spatial}
{Bovy}, J., {Rix}, H.-W., {Liu}, C., {et~al.} 2012{\natexlab{c}}, \apj, 753,
  148

\bibitem[{{Brook} {et~al.}(2007){Brook}, {Richard}, {Kawata}, {Martel}, \&
  {Gibson}}]{brook_2007}
{Brook}, C., {Richard}, S., {Kawata}, D., {Martel}, H., \& {Gibson}, B.~K.
  2007, \apj, 658, 60

\bibitem[{{Brook} {et~al.}(2004){Brook}, {Kawata}, {Gibson}, \&
  {Freeman}}]{brook_2004}
{Brook}, C.~B., {Kawata}, D., {Gibson}, B.~K., \& {Freeman}, K.~C. 2004, \apj,
  612, 894

\bibitem[{{Carlberg} {et~al.}(1985){Carlberg}, {Dawson}, {Hsu}, \&
  {Vandenberg}}]{carlberg_1985}
{Carlberg}, R.~G., {Dawson}, P.~C., {Hsu}, T., \& {Vandenberg}, D.~A. 1985,
  \apj, 294, 674

\bibitem[{{Demarque} {et~al.}(2004){Demarque}, {Woo}, {Kim}, \&
  {Yi}}]{y2_demarque_2002}
{Demarque}, P., {Woo}, J.-H., {Kim}, Y.-C., \& {Yi}, S.~K. 2004, \apjs, 155,
  667

\bibitem[{{Feltzing} {et~al.}(2001){Feltzing}, {Holmberg}, \&
  {Hurley}}]{feltzing_2011}
{Feltzing}, S., {Holmberg}, J., \& {Hurley}, J.~R. 2001, \aap, 377, 911

\bibitem[{{Fuhrmann}(1998)}]{fuhrmann_1998}
{Fuhrmann}, K. 1998, \aap, 338, 161

\bibitem[{{Gilmore} {et~al.}(2012){Gilmore}, {Randich}, {Asplund}, {Binney},
  {Bonifacio}, {Drew}, {Feltzing}, {Ferguson}, {Jeffries}, {Micela},
  {Negueruela}, {Prusti}, {Rix}, {Vallenari}, {Alfaro}, {Allende-Prieto},
  {Babusiaux}, {Bensby}, {Blomme}, {Bragaglia}, {Flaccomio}, {Fran{\c c}ois},
  {Irwin}, {Koposov}, {Korn}, {Lanzafame}, {Pancino}, {Paunzen},
  {Recio-Blanco}, {Sacco}, {Smiljanic}, {Van Eck}, \&
  {Walton}}]{ges_gilmore_2012}
{Gilmore}, G., {Randich}, S., {Asplund}, M., {et~al.} 2012, The Messenger, 147,
  25

\bibitem[{{Godwin} \& {Lynden-Bell}(1987)}]{godwin_1987}
{Godwin}, P.~J. \& {Lynden-Bell}, D. 1987, \mnras, 229, 7P

\bibitem[{{Haywood} {et~al.}(2013){Haywood}, {Di Matteo}, {Lehnert}, {Katz}, \&
  {G{\'o}mez}}]{Haywood_2013}
{Haywood}, M., {Di Matteo}, P., {Lehnert}, M.~D., {Katz}, D., \& {G{\'o}mez},
  A. 2013, \aap, 560, A109

\bibitem[{{Hernquist} \& {Quinn}(1989)}]{hernquist_1989}
{Hernquist}, L. \& {Quinn}, P.~J. 1989, in NATO Advanced Science Institutes
  (ASI) Series C, Vol. 264, NATO Advanced Science Institutes (ASI) Series C,
  ed. C.~S. {Frenk}, R.~S. {sEllis}, T.~{Shanks}, A.~R. {Heavens}, \& J.~A.
  {Peacock}, 427

\bibitem[{{Kordopatis} {et~al.}(2013){Kordopatis}, {Gilmore}, {Steinmetz},
  {Boeche}, {Seabroke}, {Siebert}, {Zwitter}, {Binney}, {de Laverny},
  {Recio-Blanco}, {Williams}, {Piffl}, {Enke}, {Roeser}, {Bijaoui}, {Wyse},
  {Freeman}, {Munari}, {Carrillo}, {Anguiano}, {Burton}, {Campbell}, {Cass},
  {Fiegert}, {Hartley}, {Parker}, {Reid}, {Ritter}, {Russell}, {Stupar},
  {Watson}, {Bienaym{\'e}}, {Bland-Hawthorn}, {Gerhard}, {Gibson}, {Grebel},
  {Helmi}, {Navarro}, {Conrad}, {Famaey}, {Faure}, {Just}, {Kos}, {Matijevi{\v
  c}}, {McMillan}, {Minchev}, {Scholz}, {Sharma}, {Siviero}, {de Boer}, \& {{\v
  Z}erjal}}]{kordopatis_2013}
{Kordopatis}, G., {Gilmore}, G., {Steinmetz}, M., {et~al.} 2013, \aj, 146, 134

\bibitem[{{Kordopatis} {et~al.}(2011){Kordopatis}, {Recio-Blanco}, {de
  Laverny}, {Gilmore}, {Hill}, {Wyse}, {Helmi}, {Bijaoui}, {Zoccali}, \&
  {Bienaym{\'e}}}]{kordopatis_2011}
{Kordopatis}, G., {Recio-Blanco}, A., {de Laverny}, P., {et~al.} 2011, \aap,
  535, A107

\bibitem[{{Lee} {et~al.}(2011{\natexlab{a}}){Lee}, {Beers}, {Allende Prieto},
  {Lai}, {Rockosi}, {Morrison}, {Johnson}, {An}, {Sivarani}, \&
  {Yanny}}]{Lee_2011a}
{Lee}, Y.~S., {Beers}, T.~C., {Allende Prieto}, C., {et~al.}
  2011{\natexlab{a}}, \aj, 141, 90

\bibitem[{{Lee} {et~al.}(2011{\natexlab{b}}){Lee}, {Beers}, {An}, {Ivezi{\'c}},
  {Just}, {Rockosi}, {Morrison}, {Johnson}, {Sch{\"o}nrich}, {Bird}, {Yanny},
  {Harding}, \& {Rocha-Pinto}}]{lee_2011b}
{Lee}, Y.~S., {Beers}, T.~C., {An}, D., {et~al.} 2011{\natexlab{b}}, \apj, 738,
  187

\bibitem[{{Lehnert} {et~al.}(2014){Lehnert}, {Di Matteo}, {Haywood}, \&
  {Snaith}}]{Lehnert_2014}
{Lehnert}, M.~D., {Di Matteo}, P., {Haywood}, M., \& {Snaith}, O.~N. 2014,
  \apjl, 789, L30

\bibitem[{{Lewis} \& {Freeman}(1989)}]{lewis_1989}
{Lewis}, J.~R. \& {Freeman}, K.~C. 1989, \aj, 97, 139

\bibitem[{{Liu} \& {van de Ven}(2012)}]{liu_2012}
{Liu}, C. \& {van de Ven}, G. 2012, \mnras, 425, 2144

\bibitem[{{Loebman} {et~al.}(2011){Loebman}, {Ro{\v s}kar}, {Debattista},
  {Ivezi{\'c}}, {Quinn}, \& {Wadsley}}]{loebman_2011}
{Loebman}, S.~R., {Ro{\v s}kar}, R., {Debattista}, V.~P., {et~al.} 2011, \apj,
  737, 8

\bibitem[{{Martig} {et~al.}(2014){Martig}, {Minchev}, \& {Flynn}}]{martig_2014}
{Martig}, M., {Minchev}, I., \& {Flynn}, C. 2014, \mnras, 443, 2452

\bibitem[{{Matteucci}(2001)}]{matteucci_2001}
{Matteucci}, F., ed. 2001, Astrophysics and Space Science Library, Vol. 253,
  {The chemical evolution of the Galaxy}

\bibitem[{{Mikolaitis} {et~al.}(2014){Mikolaitis}, {Hill}, {Recio-Blanco}, {de
  Laverny}, {Allende Prieto}, {Kordopatis}, {Tautvai{\v s}iene}, {Romano},
  {Gilmore}, {Randich}, {Feltzing}, {Micela}, {Vallenari}, {Alfaro}, {Bensby},
  {Bragaglia}, {Flaccomio}, {Lanzafame}, {Pancino}, {Smiljanic}, {Bergemann},
  {Carraro}, {Costado}, {Damiani}, {Hourihane}, {Jofr{\'e}}, {Lardo},
  {Magrini}, {Maiorca}, {Morbidelli}, {Sbordone}, {Sousa}, {Worley}, \&
  {Zaggia}}]{Sarunas}
{Mikolaitis}, {\v S}., {Hill}, V., {Recio-Blanco}, A., {et~al.} 2014, \aap,
  572, A33

\bibitem[{{Minchev} {et~al.}(2013){Minchev}, {Chiappini}, \&
  {Martig}}]{mcm_2013}
{Minchev}, I., {Chiappini}, C., \& {Martig}, M. 2013, \aap, 558, A9

\bibitem[{{Minchev} {et~al.}(2014{\natexlab{a}}){Minchev}, {Chiappini}, \&
  {Martig}}]{minchev_2014b}
{Minchev}, I., {Chiappini}, C., \& {Martig}, M. 2014{\natexlab{a}}, ArXiv
  e-prints

\bibitem[{{Minchev} {et~al.}(2014{\natexlab{b}}){Minchev}, {Chiappini},
  {Martig}, {Steinmetz}, {de Jong}, {Boeche}, {Scannapieco}, {Zwitter}, {Wyse},
  {Binney}, {Bland-Hawthorn}, {Bienaym{\'e}}, {Famaey}, {Freeman}, {Gibson},
  {Grebel}, {Gilmore}, {Helmi}, {Kordopatis}, {Lee}, {Munari}, {Navarro},
  {Parker}, {Quillen}, {Reid}, {Siebert}, {Siviero}, {Seabroke}, {Watson}, \&
  {Williams}}]{minchev_chemo_2014}
{Minchev}, I., {Chiappini}, C., {Martig}, M., {et~al.} 2014{\natexlab{b}},
  \apjl, 781, L20

\bibitem[{{Minchev} {et~al.}(2012){Minchev}, {Famaey}, {Quillen}, {Dehnen},
  {Martig}, \& {Siebert}}]{minchev_2012}
{Minchev}, I., {Famaey}, B., {Quillen}, A.~C., {et~al.} 2012, \aap, 548, A127

\bibitem[{{Nidever} {et~al.}(2014){Nidever}, {Bovy}, {Bird}, {Andrews},
  {Hayden}, {Holtzman}, {Majewski}, {Smith}, {Robin}, {Garcia Perez}, {Cunha},
  {Allende Prieto}, {Zasowski}, {Schiavon}, {Johnson}, {Weinberg}, {Feuillet},
  {Schneider}, {Shetrone}, {Sobeck}, {Garcia-Hernandez}, {Zamora}, {Rix},
  {Beers}, {Wilson}, {O'Connell}, {Minchev}, {Chiappini}, {Anders}, {Bizyaev},
  {Brewington}, {Ebelke}, {Frinchaboy}, {Ge}, {Kinemuchi}, {Malanushenko},
  {Malanushenko}, {Marchante}, {Meszaros}, {Oravetz}, {Pan}, {Simmons}, \&
  {Skrutskie}}]{nidever_2014_apogee}
{Nidever}, D.~L., {Bovy}, J., {Bird}, J.~C., {et~al.} 2014, ArXiv e-prints

\bibitem[{{Nordstr{\"o}m} {et~al.}(2004){Nordstr{\"o}m}, {Mayor}, {Andersen},
  {Holmberg}, {Pont}, {J{\o}rgensen}, {Olsen}, {Udry}, \&
  {Mowlavi}}]{nordstrom_2004}
{Nordstr{\"o}m}, B., {Mayor}, M., {Andersen}, J., {et~al.} 2004, \aap, 418, 989

\bibitem[{{Pryor} \& {Meylan}(1993)}]{pryor_1993}
{Pryor}, C. \& {Meylan}, G. 1993, in Astronomical Society of the Pacific
  Conference Series, Vol.~50, Structure and Dynamics of Globular Clusters, ed.
  S.~G. {Djorgovski} \& G.~{Meylan}, 357

\bibitem[{{Quinn} {et~al.}(1993){Quinn}, {Hernquist}, \&
  {Fullagar}}]{quinn_1993}
{Quinn}, P.~J., {Hernquist}, L., \& {Fullagar}, D.~P. 1993, \apj, 403, 74

\bibitem[{{Recio-Blanco} {et~al.}(2014){Recio-Blanco}, {de Laverny},
  {Kordopatis}, {Helmi}, {Hill}, {Gilmore}, {Wyse}, {Adibekyan}, {Randich},
  {Asplund}, {Feltzing}, {Jeffries}, {Micela}, {Vallenari}, {Alfaro}, {Allende
  Prieto}, {Bensby}, {Bragaglia}, {Flaccomio}, {Koposov}, {Korn}, {Lanzafame},
  {Pancino}, {Smiljanic}, {Jackson}, {Lewis}, {Magrini}, {Morbidelli},
  {Prisinzano}, {Sacco}, {Worley}, {Hourihane}, {Bergemann}, {Costado},
  {Heiter}, {Joffre}, {Lardo}, {Lind}, \& {Maiorca}}]{ges_disc_recio_2014}
{Recio-Blanco}, A., {de Laverny}, P., {Kordopatis}, G., {et~al.} 2014, \aap,
  567, A5

\bibitem[{{Roeser} {et~al.}(2010){Roeser}, {Demleitner}, \&
  {Schilbach}}]{roeser_2010_ppmxl}
{Roeser}, S., {Demleitner}, M., \& {Schilbach}, E. 2010, \aj, 139, 2440

\bibitem[{{Sch{\"o}nrich} \& {Binney}(2009)}]{schonrich_2009}
{Sch{\"o}nrich}, R. \& {Binney}, J. 2009, \mnras, 399, 1145

\bibitem[{{Sch{\"o}nrich} {et~al.}(2010){Sch{\"o}nrich}, {Binney}, \&
  {Dehnen}}]{schonrich_2010}
{Sch{\"o}nrich}, R., {Binney}, J., \& {Dehnen}, W. 2010, \mnras, 403, 1829

\bibitem[{{Seabroke} \& {Gilmore}(2007)}]{seabroke_2007}
{Seabroke}, G.~M. \& {Gilmore}, G. 2007, \mnras, 380, 1348

\bibitem[{{Sellwood} \& {Binney}(2002)}]{sellwood_binney_2002}
{Sellwood}, J.~A. \& {Binney}, J.~J. 2002, \mnras, 336, 785

\bibitem[{{Sharma} {et~al.}(2014){Sharma}, {Bland-Hawthorn}, {Binney},
  {Freeman}, {Steinmetz}, {Boeche}, {Bienaym{\'e}}, {Gibson}, {Gilmore},
  {Grebel}, {Helmi}, {Kordopatis}, {Munari}, {Navarro}, {Parker}, {Reid},
  {Seabroke}, {Siebert}, {Watson}, {Williams}, {Wyse}, \&
  {Zwitter}}]{sharma_2014}
{Sharma}, S., {Bland-Hawthorn}, J., {Binney}, J., {et~al.} 2014, \apj, 793, 51

\bibitem[{{Snaith} {et~al.}(2015){Snaith}, {Haywood}, {Di Matteo}, {Lehnert},
  {Combes}, {Katz}, \& {G{\'o}mez}}]{snaith_2015}
{Snaith}, O., {Haywood}, M., {Di Matteo}, P., {et~al.} 2015, \aap, 578, A87

\bibitem[{{Steinmetz} {et~al.}(2006){Steinmetz}, {Zwitter}, {Siebert},
  {Watson}, {Freeman}, {Munari}, {Campbell}, {Williams}, {Seabroke}, {Wyse},
  {Parker}, {Bienaym{\'e}}, {Roeser}, {Gibson}, {Gilmore}, {Grebel}, {Helmi},
  {Navarro}, {Burton}, {Cass}, {Dawe}, {Fiegert}, {Hartley}, {Russell},
  {Saunders}, {Enke}, {Bailin}, {Binney}, {Bland-Hawthorn}, {Boeche}, {Dehnen},
  {Eisenstein}, {Evans}, {Fiorucci}, {Fulbright}, {Gerhard}, {Jauregi}, {Kelz},
  {Mijovi{\'c}}, {Minchev}, {Parmentier}, {Pe{\~n}arrubia}, {Quillen}, {Read},
  {Ruchti}, {Scholz}, {Siviero}, {Smith}, {Sordo}, {Veltz}, {Vidrih}, {von
  Berlepsch}, {Boyle}, \& {Schilbach}}]{RAVE_2006}
{Steinmetz}, M., {Zwitter}, T., {Siebert}, A., {et~al.} 2006, \aj, 132, 1645

\bibitem[{{Vera-Ciro} {et~al.}(2014){Vera-Ciro}, {D'Onghia}, {Navarro}, \&
  {Abadi}}]{veraciro_2014}
{Vera-Ciro}, C., {D'Onghia}, E., {Navarro}, J., \& {Abadi}, M. 2014, \apj, 794,
  173

\bibitem[{{Villalobos} \& {Helmi}(2008)}]{villalobos_2008}
{Villalobos}, {\'A}. \& {Helmi}, A. 2008, \mnras, 391, 1806

\bibitem[{{Walker} {et~al.}(1996){Walker}, {Mihos}, \&
  {Hernquist}}]{walker_1996}
{Walker}, I.~R., {Mihos}, J.~C., \& {Hernquist}, L. 1996, \apj, 460, 121

\bibitem[{{Wielen}(1974)}]{wielen_1974}
{Wielen}, R. 1974, Highlights of Astronomy, 3, 395

\bibitem[{{Wielen}(1975)}]{wielen_1975}
{Wielen}, R. 1975, in La Dynamique des galaxies spirales, ed. L.~{Weliachew},
  357

\bibitem[{{Wielen}(1977)}]{wielen_1977}
{Wielen}, R. 1977, \aap, 60, 263

\bibitem[{{Yanny} {et~al.}(2009){Yanny}, {Rockosi}, {Newberg}, {Knapp},
  {Adelman-McCarthy}, {Alcorn}, {Allam}, {Allende Prieto}, {An}, {Anderson},
  {Anderson}, {Bailer-Jones}, {Bastian}, {Beers}, {Bell}, {Belokurov},
  {Bizyaev}, {Blythe}, {Bochanski}, {Boroski}, {Brinchmann}, {Brinkmann},
  {Brewington}, {Carey}, {Cudworth}, {Evans}, {Evans}, {Gates}, {G{\"a}nsicke},
  {Gillespie}, {Gilmore}, {Nebot Gomez-Moran}, {Grebel}, {Greenwell}, {Gunn},
  {Jordan}, {Jordan}, {Harding}, {Harris}, {Hendry}, {Holder}, {Ivans},
  {Ivezi{\v c}}, {Jester}, {Johnson}, {Kent}, {Kleinman}, {Kniazev},
  {Krzesinski}, {Kron}, {Kuropatkin}, {Lebedeva}, {Lee}, {French Leger},
  {L{\'e}pine}, {Levine}, {Lin}, {Long}, {Loomis}, {Lupton}, {Malanushenko},
  {Malanushenko}, {Margon}, {Martinez-Delgado}, {McGehee}, {Monet}, {Morrison},
  {Munn}, {Neilsen}, {Nitta}, {Norris}, {Oravetz}, {Owen}, {Padmanabhan},
  {Pan}, {Peterson}, {Pier}, {Platson}, {Re Fiorentin}, {Richards}, {Rix},
  {Schlegel}, {Schneider}, {Schreiber}, {Schwope}, {Sibley}, {Simmons},
  {Snedden}, {Allyn Smith}, {Stark}, {Stauffer}, {Steinmetz}, {Stoughton},
  {SubbaRao}, {Szalay}, {Szkody}, {Thakar}, {Sivarani}, {Tucker}, {Uomoto},
  {Vanden Berk}, {Vidrih}, {Wadadekar}, {Watters}, {Wilhelm}, {Wyse}, {Yarger},
  \& {Zucker}}]{SEGUE_2009}
{Yanny}, B., {Rockosi}, C., {Newberg}, H.~J., {et~al.} 2009, \aj, 137, 4377

\end{thebibliography}

\end{document}